\documentclass[notitlepage,nofootinbib,preprintnumbers,amssymb,superscriptaddress]{revtex4-2}
\usepackage{amsfonts,amssymb,mathtools,graphicx,color,bm,booktabs,siunitx}
\definecolor{ultramarine}{rgb}{0.07, 0.04, 0.56}
\definecolor{cadmiumgreen}{rgb}{0.0, 0.42, 0.24}
\definecolor{indigo(dye)}{rgb}{0.0, 0.25, 0.42}
\usepackage[linktocpage=true]{hyperref}
\hypersetup{
colorlinks=true,
citecolor=ultramarine,
linkcolor=cadmiumgreen,
urlcolor=indigo(dye),
}

\let\originalleft\left
\let\originalright\right
\renewcommand{\left}{\mathopen{}\mathclose\bgroup\originalleft}
\renewcommand{\right}{\aftergroup\egroup\originalright}
\usepackage{autobreak}
\newcommand{\D}{{\rm d}}
\newcommand{\fr}[2]{\frac{#1}{#2}}
\newcommand{\pa}{\partial}
\newcommand{\ti}{\tilde}
\newcommand{\na}{\nabla}
\newcommand{\bra}[1]{\left( #1 \right)}  
\newcommand{\brb}[1]{\left[ #1 \right]}  
  
\newcommand{\be}{\begin{equation}}  
\newcommand{\ee}{\end{equation}}
\newcommand{\bem}{\begin{bmatrix}}
\newcommand{\eem}{\end{bmatrix}}

\newcommand{\vae}{\varepsilon}

\newcommand{\mn}{{\mu \nu}}

\newcommand{\mP}{\mathcal{P}}

\begin{document}

\preprint{IPMU25-0055}

\title{Scalar quasinormal modes of rotating black holes in parity-violating gravity}

\author{Hiroaki W.~H.~Tahara}
\affiliation{Department of Physics, Tokyo Metropolitan University, 1-1 Minami-Osawa, Hachioji, Tokyo 192-0397, Japan}

\author{Hayato Motohashi}
\affiliation{Department of Physics, Tokyo Metropolitan University, 1-1 Minami-Osawa, Hachioji, Tokyo 192-0397, Japan}
\affiliation{Yukawa Institute for Theoretical Physics, Kyoto University, 606-8502, Kyoto, Japan}

\author{Kazufumi Takahashi}
\affiliation{Department of Physics, College of Humanities and Sciences, Nihon University, Tokyo 156-8550, Japan}
\affiliation{Yukawa Institute for Theoretical Physics, Kyoto University, 606-8502, Kyoto, Japan}

\author{Vicharit Yingcharoenrat}
\affiliation{High Energy Physics Research Unit, Department of Physics, Faculty of Science, Chulalongkorn University, Pathumwan, Bangkok 10330, Thailand}
\affiliation{Kavli Institute for the Physics and Mathematics of the Universe (WPI), The University of Tokyo, Kashiwa, Chiba 277-8583, Japan}

\begin{abstract}
Recently, an exact rotating black hole solution in a parity-violating theory of gravity was obtained via a conformal transformation of the Kerr solution in general relativity, with parity-violating effects encoded in the conformal factor.
We study the quasinormal modes (QNMs) of a test scalar field minimally coupled to gravity on this conformal Kerr background, treating the parity-violating effects perturbatively while allowing for arbitrary black hole spin, from the non-rotating case to the near-extremal regime.
For low spin, we derive a perturbative formula for the QNM frequencies that includes the leading-order parity-violating correction.
For high spin, particularly in the near-extremal regime, we find sizable deviations from the Kerr QNM frequencies.
Our results point to a new avenue for probing parity-violating physics in the strong-gravity regime through black hole QNMs.
\end{abstract}

\maketitle

\section{Introduction}\label{sec:intro}

Parity violation is a well-established feature of fundamental interactions, manifesting itself in a wide range of physical phenomena.
In the 1950s, parity symmetry was shown to be violated in the weak interaction through observations of beta decay~\cite{Lee:1956qn,Wu:1957my}, and this remains the only such violation observed experimentally in nature.
Parity violation may also arise in physics beyond the Standard Model or in the gravitational sector through astrophysical or cosmological processes.
For example, parity-violating effects in the early Universe and their imprints on the cosmic microwave background have been extensively studied; see, e.g., Refs.~\cite{Lue:1998mq,Alexander:2004us,Li:2006ss,Saito:2007kt,Sorbo:2011rz,Soda:2011am,Kaufman:2014rpa,Bartolo:2017szm,Bartolo:2018elp,Minami:2020odp,Wang:2021gqm,Komatsu:2022nvu,Caravano:2022epk,Philcox:2023ffy,Figueroa:2023oxc,Philcox:2023ypl}.
See also Refs.~\cite{Coulton:2023oug,Kurita:2025hmp,Gao:2025yqd,Azyzy:2025zif} for recent studies of large-scale structure in the presence of parity violation.
Additionally, attempts to detect parity violation in the Universe through gravitational-wave observations have been explored in Refs.~\cite{Seto:2007tn,Yunes:2010yf,Cook:2011hg,Yagi:2017zhb,Alexander:2017jmt,Nishizawa:2018srh,Zhao:2019xmm,Qiao:2019hkz,Yamada:2020zvt,Perkins:2020tra,Perkins:2021mhb,Gong:2021jgg,Okounkova:2021xjv,Qiao:2022mln,Jenks:2023pmk}.

In the context of gravitational parity violation, there have been extensive studies of Chern-Simons gravity with either a dynamical or a nondynamical scalar field~\cite{Jackiw:2003pm,Smith:2007jm,Alexander:2009tp}.
Perturbations about a static and spherically symmetric black hole (BH) background have been studied in Refs.~\cite{Yunes:2007ss,Cardoso:2009pk,Molina:2010fb,Garfinkle:2010zx,Pani:2011xj,Motohashi:2011pw,Motohashi:2011ds,Kimura:2018nxk,Macedo:2018txb,Wagle:2021tam}.
There have also been studies of rotating BH solutions in Chern-Simons gravity~\cite{Konno:2007ze,Yunes:2009hc,Konno:2009kg,Pani:2011gy,Yagi:2012ya,Stein:2014xba,Berti:2015itd,McNees:2015srl,Maselli:2017kic,Berti:2018cxi,Delsate:2018ome,Cunha:2018uzc,Srivastava:2021imr} either numerically or analytically using slow-rotation or near-extremal approximations, as well as investigations of their perturbations~\cite{Wagle:2023fwl,Chung:2025gyg,Li:2025fci}.
As demonstrated in Refs.~\cite{Motohashi:2011pw,Motohashi:2011ds}, Chern-Simons gravity typically involves ghost degrees of freedom, and the theory should therefore be regarded as an effective field theory, with a cutoff below the mass scales of the ghost modes.
Meanwhile, in Ref.~\cite{Tahara:2023pyg}, an exact rotating BH solution was obtained by performing a conformal transformation on the Kerr solution in general relativity (GR), with the conformal factor given as a function of the Chern-Simons scalar.
Such a conformal transformation was shown to be invertible in general~\cite{Takahashi:2022mew}, which ensures that the resulting (parity-violating) gravitational theory is ghost-free~\cite{Domenech:2015tca,Takahashi:2017zgr}, and that the metric generated by the transformation is a solution of the new gravitational theory.
A remarkable feature of this BH background is that the metric no longer respects equatorial symmetry, in sharp contrast to the Kerr BH in GR.
Therefore, this parity-violating BH background (i.e., the \emph{conformal Kerr} BH) provides a useful setting for exploring the physical implications of parity violation in gravity.
As an example, the effects of parity violation on the geodesic motion of a test particle were studied in Ref.~\cite{Tahara:2023pyg}.

Motivated by these developments, the main goal of the present paper is to study the spectrum of quasinormal modes (QNMs) of the conformal Kerr BH.
For simplicity, we focus on a test scalar field minimally coupled to gravity.
Scalar QNMs on Schwarzschild and Kerr backgrounds in GR have been extensively studied in the literature (see, e.g., Refs.~\cite{Berti:2005ys,Berti:2006wq,Berti:2009kk}), and therefore provide valuable benchmarks for comparison.
As we shall clarify later, the effects of parity violation are controlled by two parameters: the BH spin and a dimensionless parameter, denoted by $\hat{\alpha}$, which characterizes the dependence of the conformal factor on the Chern-Simons scalar.
We focus on regimes in which the parity-violating effects can be treated perturbatively, while allowing for arbitrary BH spin, from the non-rotating case to the near-extremal regime.
Based on this setup, for low spin, we derive a perturbative formula for the QNM frequencies that incorporates the leading-order parity-violating correction.
For high spin, particularly in the near-extremal regime, we find sizable deviations from the Kerr QNM frequencies.

The rest of this paper is organized as follows.
In Sec.~\ref{sec:conformal_Kerr}, we briefly review the conformal Kerr solution, including the conditions required for the invertibility of the conformal transformation.
In Sec.~\ref{sec:QNM}, we describe our approximation schemes as well as the computational methods used to obtain the QNMs, and we present the corresponding results in Sec.~\ref{sec:res}.
Finally, we conclude and discuss future directions in Sec.~\ref{sec:conc}.

\section{Conformal Kerr background}\label{sec:conformal_Kerr}

Let us briefly review the conformal Kerr solution constructed in Ref.~\cite{Tahara:2023pyg}.
The basic idea is that, given a solution in a gravitational theory, an invertible transformation can map it to a solution in the transformed theory.
The invertibility guarantees that the number of physical degrees of freedom is preserved under the transformation~\cite{Domenech:2015tca,Takahashi:2017zgr} and that solutions are mapped to one another in a one-to-one manner.
In Ref.~\cite{Tahara:2023pyg}, the seed solution was chosen to be the Kerr solution in GR, and a conformal transformation of the following form was used to obtain an exact rotating BH solution in parity-violating gravity:
    \begin{align}
    g_{\mu\nu}\to \bar{g}_{\mu\nu}[g]=\Omega({\cal P})g_{\mu\nu}\;, \qquad
    {\cal P}\coloneqq \fr{1}{2}\vae^{\alpha\beta\gamma\delta}R^{\mu\nu}{}_{\alpha\beta}R_{\mu\nu\gamma\delta} = \frac{1}{2}\vae^{\alpha\beta\gamma\delta} \mathcal{W}^{\mu\nu}{}_{\alpha\beta}\mathcal{W}_{\mu\nu\gamma\delta}\;,
    \label{conformal_trnsf}
    \end{align}
where $R_{\mu\nu\alpha\beta}$ denotes the Riemann tensor, $\mathcal{W}_{\mu\nu\alpha\beta}$ the Weyl tensor, and the conformal factor~$\Omega$ is a function of the Chern-Simons (or Pontryagin) scalar~${\cal P}$.
It should be noted that $\mP$ is a pseudo-scalar, as reflected in the presence of the totally antisymmetric tensor~$\vae^{\alpha\beta\gamma\delta}$.
As clarified in Refs.~\cite{Takahashi:2022mew,Tahara:2023pyg}, the transformation~\eqref{conformal_trnsf} is invertible provided that the conformal factor~$\Omega({\cal P})$ satisfies the following condition:
    \begin{align}
    1-\fr{2{\cal P}}{\Omega}\fr{\D\Omega}{\D{\cal P}}\ne 0 \;. \label{invertible_cond}
    \end{align}
Although one may more generally allow the conformal factor to depend on additional scalar invariants constructed from the Weyl tensor, while maintaining invertibility as discussed in Ref.~\cite{Tahara:2023pyg}, we restrict our attention to the form~\eqref{conformal_trnsf} for simplicity.

To further discuss the conformal Kerr metric~$\bar g_{\mu\nu}$, defined as
    \begin{align}
    \bar{g}_\mn=\Omega({\cal P})g_\mn\quad
    (\text{$g_\mn$: Kerr metric})\;,
    \label{conformal_Kerr}
    \end{align}
we first introduce the Kerr metric in the Boyer-Lindquist coordinates as~\cite{Carter:1970ea,Carter:1973rla}
    \be\label{eq:Kerr}
    {g}_{\mu\nu}{\rm d} x^\mu {\rm d} x^\nu = 
    - \frac{\Delta}{\rho^2} ({\rm d} t-a\sin^2\theta {\rm d}\varphi)^2 
    + \frac{\rho^2}{\Delta}{\rm d} r^2
    + {\rho^2}{\rm d}\theta^2
    + \frac{\sin^2\theta}{\rho^2} \left[ a{\rm d} t-(r^2+a^2) {\rm d}\varphi \right]^2\;,
    \ee
where we have defined
    \be
    \rho^2 \coloneqq  r^2+a^2\cos^2\theta\;, \qquad
    \Delta \coloneqq  r^2+a^2 - 2M r\;, 
    \ee
with $a$ being the angular momentum per unit mass and $M\,(>0)$ being a parameter of length dimension corresponding to the BH mass.
For $0<|a|<M$, the Kerr metric has two horizons at $r=r_\pm\coloneqq M\pm\sqrt{M^2-a^2}$, which satisfy $\Delta(r_\pm)=0$, with the outer horizon~$r=r_+$ corresponding to the event horizon. 
For later convenience, we denote $r_+$ by $r_\mathrm{H}$.
Using the Kerr metric~\eqref{eq:Kerr}, the Chern-Simons scalar~$\mP$ reads
    \be
    \mP=-\fr{96aM^2r\cos\theta(r^2-3a^2\cos^2\theta)(3r^2-a^2\cos^2\theta)}{\rho^{12}}\;. \label{eq:CS_term}
    \ee
It is evident that $\mP$ vanishes in the limit~$a\to 0$, indicating the absence of parity violation for non-spinning BHs.
For non-vanishing $a$, Eq.~\eqref{eq:CS_term} shows that $\mP$ decays asymptotically as $r\to\infty$ ($\mP\sim r^{-7}$), while developing a nontrivial profile in the vicinity of the BH horizon.
Note also that $\mP$ is a bounded function in the spacetime region outside the BH horizon.
In particular, the maximum value of $|\mP|$ is given by $|\mP|_{\rm max}=M^{-4}\brb{9|a|/(4M)+{\cal O}(a^2/M^2)}$ for low spin, and $|\mP|_{\rm max}\simeq 40.1836M^{-4}(M/r_{\rm H})^6$ for high spin (see Appendix~A of Ref.~\cite{Tahara:2023pyg} for a more detailed discussion).
Moreover, a reflection with respect to the equatorial plane~$\theta=\pi/2$ (i.e., $\theta\to\pi-\theta$) flips the sign of $\mP$, implying that the conformal Kerr metric~\eqref{conformal_Kerr} does not respect this reflection symmetry unless the conformal factor satisfies $\Omega(-\mP)=\Omega(\mP)$.

In what follows, in addition to the invertibility condition~\eqref{invertible_cond}, we assume that $\Omega\to 1$ as $\mP\to 0$, so that the conformal Kerr metric~\eqref{conformal_Kerr} asymptotically approaches the Kerr metric as $r\to\infty$.
Apart from this requirement, the conformal factor~$\Omega$ is allowed to be an otherwise arbitrary function of $\mP$.
We therefore write the Taylor expansion of $\Omega$ about $\mP=0$ as
    \begin{align}
    \Omega(\mP)=1+\frac{\alpha\mP}{\Lambda^4}+{\cal O}(\Lambda^{-8}\mP^2)\;, \qquad
    \alpha\coloneqq \left.\Lambda^4\frac{\D\Omega}{\D\mP}\right|_{\mP=0}\;, \label{Omega_Taylor_Lambda}
    \end{align}
where $\Lambda$ is an energy scale that is a priori independent of the BH mass parameter~$M$.\footnote{As a concrete example, the following form of the conformal factor was studied in Ref.~\cite{Tahara:2023pyg}:
    \begin{align}
    \Omega(\mP)=1+\tanh\bra{\frac{\alpha\mP}{\Lambda^4}}\;, \nonumber
    \end{align}
for which one can show that the invertibility condition~\eqref{invertible_cond} is satisfied for any finite value of $\mP$.
In the present paper, however, we do not restrict ourselves to this particular choice of $\Omega$.}
When applying the conformal transformation to the Kerr background, it is convenient to recast Eq.~\eqref{Omega_Taylor_Lambda} as
    \begin{align}
    \Omega(\mP)\simeq 1+\hat{\alpha}M^4\mP\;, \qquad
    \hat{\alpha}\coloneqq \frac{\alpha}{M^4\Lambda^4}\;. \label{Omega_Taylor}
    \end{align}
Below, we focus on regimes in which the parity-violating effects can be treated perturbatively and retain only the leading-order contributions, which are characterized by the dimensionless parameter~$\hat{\alpha}$.
Note that, at leading order in the parity-violating effects, the left-hand side of the invertibility condition~\eqref{invertible_cond} reduces to $1-2\hat{\alpha}M^4\mP$, which is always positive in the perturbative regime~$|\hat{\alpha}M^4\mP|\ll 1$:
    \begin{align}
    1-\fr{2{\cal P}}{\Omega}\fr{\D\Omega}{\D{\cal P}}
    \simeq 1-2\hat{\alpha}M^4\mP>0\;. \label{invertible_cond_leading}
    \end{align}
We emphasize that the effects of parity violation are controlled by two parameters, namely the dimensionless parameter~$\hat{\alpha}$ introduced above and the BH spin~$a$.
As a result, the condition $|\hat{\alpha}M^4\mP|\ll 1$ does not necessarily imply that both $\hat{\alpha}$ and $a$ are small.
For low spin, where $|M^4\mP|$ is at most ${\cal O}(|a|/M)$, the parameter~$\hat{\alpha}$ can be of order unity.
Conversely, for sufficiently small $\hat{\alpha}$, the BH spin can be near extremal.\footnote{For instance, when $a/M=0.99$, the approximate maximal value of $|M^4\mP|$ is $22.6826$ [see the discussion below Eq.~\eqref{eq:CS_term}], and the condition~$|\hat{\alpha}M^4\mP|\ll 1$ is satisfied provided that $|\hat{\alpha}|\ll 0.044$.
In Sec.~\ref{ssec:QNMs_small_alpha}, we nevertheless consider values of $\hat{\alpha}$ up to $0.02$, which approach the edge of the perturbative regime, in order to maximize the parity-violating effects.
Note in passing that this choice remains consistent with the invertibility condition~\eqref{invertible_cond_leading}.
\label{footnote2}}

\section{Quasinormal modes of a test scalar field}\label{sec:QNM}

In this section, we study the QNMs of a test scalar field~$\phi$ on the conformal Kerr background.
We assume that the scalar field is minimally coupled to the conformally transformed metric~$\bar{g}_\mn$ and therefore satisfies the following Klein-Gordon equation:
    \be
    \bar{\Box}\phi - {\mu}^2\phi
    =\fr{1}{\Omega}\Box\phi+\fr{1}{\Omega^2}\na^\alpha\Omega\na_\alpha\phi - {\mu}^2\phi
    =0\;,
    \label{KG_conformal_Kerr}
    \ee
where $\mu$ is a mass parameter, and $\bar{\Box}$ and $\Box$ denote the d'Alembert operators associated with the conformal Kerr metric~$\bar{g}_{\mu\nu}$ and the Kerr metric~$g_{\mu\nu}$, respectively.
By redefining the scalar field as
    \be
    \tilde{\phi}\coloneqq \Omega^{1/2}\phi\;,
    \ee
Eq.~\eqref{KG_conformal_Kerr} can be recast in the form of a Klein-Gordon equation on the Kerr background with an effective mass term,
    \be
    \Box\ti{\phi}-\ti{\mu}^2\ti{\phi}=0\;, \qquad
    \ti{\mu}^2\coloneqq \Omega\mu^2 + \Omega^{-1/2}\Box(\Omega^{1/2})\;.
    \label{KG_mod}
    \ee
Since $\Omega$ is a function of the Chern-Simons scalar given in Eq.~\eqref{eq:CS_term}, the effective mass squared~$\ti{\mu}^2$ depends on $(r,\theta)$ even when $\mu^2=0$.
Also, $\ti{\mu}^2$ approaches $\mu^2$ at spatial infinity, since we assume that $\Omega\to 1$ as $\mP\to 0$.
In the perturbative regime~$|\hat{\alpha} M^4 \mP| \ll 1$, one can employ Eq.~\eqref{Omega_Taylor} to obtain
    \begin{align}
    \ti{\mu}^2 - \Omega\mu^2 
    \simeq \hat{\alpha}a\cos\theta\frac{288M^6(20r-49M)}{r^{10}}\;,
    \label{eff_mass}
    \end{align}
at leading order in the parity-violating effects.
For the Kerr background, it is well known that the Klein-Gordon equation can be solved using the method of separation of variables.
In particular, after separating the time and azimuthal dependence as ${\rm e}^{-{\rm i}\omega t+{\rm i}m\varphi}$, the scalar field can be expanded in terms of spheroidal harmonics~$S_{\ell m}(\theta)$, with each mode function evolving independently.
However, as far as we have investigated, such a separation is no longer possible for the conformal Kerr background due to the presence of the effective mass term.
As we shall see, this leads to couplings between modes with different values of $\ell$.

In what follows, we focus on the case of a massless scalar on the conformal Kerr background, $\mu^2=0$, for simplicity, although the extension to the massive scalar case with $\mu^2\neq 0$ would be straightforward.
As mentioned earlier, we restrict our attention to regimes in which the parity-violating effects can be treated perturbatively, which is the case when either the BH spin~$a/M$ or the parameter~$\hat{\alpha}$ is small.
We study the low-spin case in Sec.~\ref{ssec:low_spin} and the small-$\hat{\alpha}$ case in Sec.~\ref{ssec:small_coupling_constant}.
We use different but complementary methods for computing the QNMs, each suited to one of these two cases: the matrix-valued version of Leaver's method (see, e.g., Refs.~\cite{Pani:2013pma,Nomura:2021efi}) and the spectral method~\cite{Chung:2023zdq,Chung:2023wkd}, respectively.
Note that these two cases are not mutually exclusive, and we shall see that the results are consistent when both $a$ and $\hat{\alpha}$ are small.

\subsection{Low spin}\label{ssec:low_spin}

In this section, we study the scalar QNM frequencies in the low-spin regime, i.e., $|a|/M\ll 1$, while allowing $\hat{\alpha}$ to be of order unity.
We begin by deriving the basic equation used to compute the QNMs.
As in the case of the Kerr background, we express the scalar field~$\ti{\phi}$ as (see, e.g., Ref.~\cite{Ghosh:2023etd})
 \begin{align}
    \tilde{\phi}
    =\int{\rm d}\omega\sum_{\ell,m}{\rm e}^{-{\rm i}\omega t+{\rm i}m\varphi}\frac{Z_{\ell m}(r)}{\sqrt{r^2+a^2}}S_{\ell m}(\theta)\;,
    \label{phi_ansatz}
    \end{align}
where $S_{\ell m}(\theta)$ denotes the scalar spheroidal harmonics (more precisely, divided by ${\rm e}^{{\rm i}m\varphi}$), satisfying the following differential equation~\cite{BHPToolkit}:
    \begin{align}
    \frac{1}{\sin\theta}\frac{\D}{\D\theta}\bra{\sin\theta\frac{\D S_{\ell m}}{\D\theta}}+\bra{-a^2\omega^2\sin^2\theta-\frac{m^2}{\sin^2\theta}+2ma\omega+\lambda_{\ell m}}S_{\ell m}=0\;,
    \end{align}
with $\lambda_{\ell m}$ being a separation constant. 
We note that $S_{\ell m}$ form a complete basis for functions of $\theta$ and satisfy the following bi-orthogonality relation~\cite{London:2020uva}:
    \begin{align}
    2\pi\int_{-1}^{1}\D(\cos\theta)\,S_{\ell' m}\bar{S}_{\ell m}^*=\delta_{\ell\ell'}\;,
    \label{orthogonality}
    \end{align}
where $\bar{S}_{\ell m}$ is the adjoint-spheroidal harmonics and an asterisk denotes complex conjugation.
Substituting the ansatz~\eqref{phi_ansatz} into the modified Klein-Gordon equation~\eqref{KG_mod} with $\mu^2=0$, we obtain
    \begin{align}
    \sum_{\ell'}S_{\ell' m}\brb{\fr{\D^2}{\D r_*^2}+U_{\ell' m}^{(0)}-\frac{\rho^2\Delta}{(r^2+a^2)^2}\ti{\mu}^2}Z_{\ell' m}=0\;,
    \label{KG_mod2}
    \end{align}
where $r_*$ denotes the tortoise coordinate, which satisfies $r_*\to -\infty$ as $r\to r_{\rm H}$ and $r_*\to \infty$ as $r\to \infty$, defined by
    \begin{align}
    \fr{\D r_*}{\D r}=\frac{r^2+a^2}{\Delta}\;,
    \end{align}
and $U_{\ell m}^{(0)}(r)$ is the effective potential for the Kerr background, given by
    \begin{align}\label{eq:potential_Kerr}
    U_{\ell m}^{(0)}(r)=\frac{[(r^2+a^2)\omega-m a]^2-\lambda_{\ell m}\Delta}{(r^2+a^2)^2}-\frac{1}{\sqrt{r^2+a^2}}\fr{\D^2}{\D r_*^2}\sqrt{r^2+a^2}\;.
    \end{align}
In the limit~$a\to 0$, the spheroidal harmonics~$S_{\ell m}(\theta)$ reduce to the associated Legendre polynomials~$P_\ell^{|m|}(\cos\theta)$, i.e.,
    \begin{align}
    S_{\ell m}(\theta)\to \sqrt{\frac{2\ell+1}{4\pi}\frac{(\ell-|m|)!}{(\ell+|m|)!}}\,P_\ell^{|m|}(\cos\theta) \;,
    \label{Slm_a=0}
    \end{align}
and the separation constant satisfies $\lambda_{\ell m}\to \ell(\ell+1)$.
In this limit, the potential~\eqref{eq:potential_Kerr} reduces to
    \begin{align}
    U_{\ell m}^{(0)}(r)\to \omega^2-\bra{1-\frac{2M}{r}}\brb{\frac{\ell(\ell+1)}{r^2}+\frac{2M}{r^3}}\eqqcolon \omega^2-V_{\rm RW}(r)\;,
    \end{align}
where $V_{\rm RW}(r)$ denotes the Regge-Wheeler potential for a spin-zero field.
Note also that, using the series expansion of $\lambda_{\ell m}$~\cite{BHPToolkit},
    \begin{align}
    \lambda_{\ell m}=\ell(\ell+1)-2ma\omega
    +\frac{2[\ell(\ell+1)+m^2-1]}{(2\ell-1)(2\ell+3)}a^2\omega^2
    +{\cal O}(a^3\omega^3)\;,
    \end{align}
we obtain
    \begin{align}
    U_{\ell m}^{(0)}&\simeq \omega^2-V_{\rm RW}(r)-\frac{4mMa\omega}{r^3}
    -\frac{2a^2\omega^2[\ell(\ell+1)+m^2-1]}{(2\ell-1)(2\ell+3)r^2}\bra{1-\frac{2M}{r}} \nonumber \\
    &\quad +\frac{a^2\ell(\ell+1)}{r^4}\bra{1-\frac{4M}{r}}+\frac{m^2a^2}{r^4}-\frac{a^2}{r^4}\bra{1-\frac{12M}{r}+\frac{24M^2}{r^2}}\;. \label{Ulm}
    \end{align}

Multiplying both sides of Eq.~\eqref{KG_mod2} by $\bar{S}_{\ell m}^*$ and integrating over $\theta$, we obtain
    \begin{align}
    \brb{\fr{\D^2}{\D r_*^2}+U_{\ell m}^{(0)}(r)}Z_{\ell m}+\sum_{\ell'}u^m_{\ell\ell'}Z_{\ell'm}=0\;, \qquad
    u^m_{\ell\ell'}\coloneqq -\frac{\Delta}{(r^2+a^2)^2}\cdot 2\pi\int_{-1}^{1}\D(\cos\theta)\,S_{\ell' m}\bar{S}_{\ell m}^*\rho^2\ti{\mu}^2\;,
    \label{master_eq}
    \end{align}
where we have used the bi-orthogonality relation~\eqref{orthogonality}.
It should be noted that $u^m_{\ell\ell'}$ vanishes both at the BH horizon, $r=r_\mathrm{H}$ where $\Delta=0$, and at spatial infinity.
In addition, the potential~$U_{\ell m}^{(0)}(r)$ behaves as 
    \begin{align}
    U_{\ell m}^{(0)}(r)\to
    \left\{\begin{array}{ll}
    (\omega-m\omega_{\rm H})^2 \;, & \qquad r\to r_\mathrm{H}\;, \\
    \omega^2 \;, & \qquad r\to \infty\;,
    \end{array}\right.
    \end{align}
with $\omega_{\rm H}\coloneqq a/(r_\mathrm{H}^2+a^2)=a/(2Mr_\mathrm{H})$.
Therefore, the QNM boundary conditions for $Z_{\ell m}(r)$ can be written as
    \begin{align}
    Z_{\ell m}(r)\sim
    \left\{\begin{array}{ll}
    {\rm e}^{-{\rm i}(\omega-m\omega_{\rm H})r_*} \;, & \qquad r\to r_\mathrm{H}\;, \\
    {\rm e}^{{\rm i}\omega r_*} \;, & \qquad r\to \infty\;,
    \end{array}\right.
    \label{QNM_BC}
    \end{align}
which correspond to purely ingoing and outgoing boundary conditions at $r = r_{\rm H}$ and at infinity, respectively.
From Eq.~\eqref{master_eq}, we see that infinitely many modes with different multipole indices are coupled with each other.
Indeed, if one focuses on a mode with given $(\ell,m)$, it couples to modes with $(\ell',m)$ such that $\ell'=\ell\pm1, \ell\pm3, \ell\pm5, \cdots$ ($\ell'\ge |m|$).
This selection rule follows from the fact that the effective mass squared~$\tilde{\mu}^2$ in the integrand of Eq.~\eqref{master_eq} is odd with respect to $\cos\theta$ [see Eq.~\eqref{eff_mass} with $\mu^2=0$].
However, at leading order in the parity-violating effects, only a finite number of modes are coupled with each other, as we shall see below.
By use of Eq.~\eqref{eff_mass} with $\mu^2=0$, the coefficient~$u^m_{\ell\ell'}$ in Eq.~\eqref{master_eq} can be evaluated as
    \begin{align}
    u^m_{\ell\ell'}
    &\simeq -A^m_{\ell\ell'}\bra{1-\frac{2M}{r}}\frac{288M^7(20r-49M)}{r^{10}}\frac{\hat{\alpha}a}{M}\;,
    \label{mode_coupling}
    \end{align}
at leading order in $a/M$, where $A^m_{\ell\ell'}$ is a constant given by
    \begin{align}
    A^m_{\ell\ell'}\coloneqq 2\pi C_{\ell'}^m C_\ell^m\int_{-1}^{1}\D\chi\,P_{\ell'}^{|m|} (\chi)P_\ell^{|m|}(\chi)\chi\;, \qquad
    C_\ell^m\coloneqq \sqrt{\frac{2\ell+1}{4\pi}\frac{(\ell-|m|)!}{(\ell+|m|)!}}\;,
    \end{align}
with $\chi=\cos\theta$.
Note in passing that, in deriving Eq.~\eqref{mode_coupling}, we did not assume $|\hat{\alpha}|\ll 1$: extracting the leading-order term in $a/M$ automatically captures only the term linear in $\hat\alpha$.
Note also that $A^m_{\ell\ell'}$ is non-vanishing only when $\ell'=\ell\pm 1$.
Thus, if one is interested in a mode with $(\ell,m)$, it is sufficient to take into account couplings to modes with $(\ell\pm 1,m)$.
Therefore, at leading orders in the BH spin, Eq.~\eqref{master_eq} can be written in the following form:
    \begin{align}
    \brb{\mathbb{I}_{3}\fr{\D^2}{\D r_*^2}+\mathbb{U}_{\ell m}(r)}
    \begin{pmatrix}
    Z_{\ell-1,m}\\
    Z_{\ell m}\\
    Z_{\ell+1,m}
    \end{pmatrix}=0\;, \qquad
    \mathbb{U}_{\ell m}\coloneqq
    \begin{pmatrix}
    U_{\ell-1,m}^{(0)}&u^m_{\ell-1,\ell}&0\\
    u^m_{\ell,\ell-1}&U_{\ell,m}^{(0)}&u^m_{\ell,\ell+1}\\
    0&u^m_{\ell+1,\ell}&U_{\ell+1,m}^{(0)}
    \end{pmatrix}\;,
    \label{master_eq2}
    \end{align}
with $\mathbb{I}_{3}$ denoting the $3\times 3$ identity matrix.
The off-diagonal components of the matrix-valued effective potential~$\mathbb{U}_{\ell m}$, given explicitly in Eq.~\eqref{mode_coupling}, encode the parity-violating effects.
Note that, for modes with $\ell=|m|$, the variable~$Z_{\ell-1,m}$ is absent, and Eq.~\eqref{master_eq2} should be understood as a reduced system obtained by discarding the first row and first column of the matrix-valued operator.

We have now obtained a set of coupled second-order ordinary differential equations.
To compute the QNMs, we apply the matrix-valued version of Leaver's method (see, e.g., Refs.~\cite{Pani:2013pma,Nomura:2021efi}).
Specifically, we solve Eq.~\eqref{master_eq2} subject to the boundary conditions~\eqref{QNM_BC} by assuming a Frobenius series expansion.
This leads to a many-term recurrence relation for the $\omega$-dependent, matrix-valued series coefficients.
The recurrence relation can be reduced to a three-term one by Gaussian elimination and subsequently recast as a matrix-valued continued-fraction equation.
By truncating the continued fraction at a finite order, we obtain an algebraic equation for~$\omega$, schematically written as ${\cal A}(\omega;a,\hat{\alpha})=0$, whose solutions yield the QNM frequencies.\footnote{Rather than simply truncating the continued fraction, one can obtain a more accurate approximation by replacing the truncated tail with an estimate based on the asymptotic behavior of the higher-order coefficients of the Frobenius series.
This procedure, known as Nollert’s improvement~\cite{Nollert:1993zz,Zhidenko:2006rs}, is implemented in our numerical computation of the QNMs.}
The truncation order is chosen to be sufficiently large, and in practice is set to 50, such that the QNM frequencies of interest are stable under variations of the truncation order.

Let us now study the behavior of the QNM frequencies in the low-spin regime.
To this end, it is useful to first examine the structure of the QNM-determining equation, ${\cal A}(\omega;a,\hat{\alpha})=0$, and in particular how the dependence on~$\hat{\alpha}$ enters.
As seen from Eq.~\eqref{master_eq2}, the off-diagonal components of the matrix~$\mathbb{U}_{\ell m}$ are proportional to $\hat{\alpha}a/M$ at leading order in the BH spin.
Since products of these off-diagonal terms enter the equation~${\cal A}(\omega;a,\hat{\alpha})=0$, deviations from the Kerr QNM frequencies arise only at ${\cal O}(\hat{\alpha}^2a^2/M^2)$.
As a consequence, the QNM frequency admits the following series expansion in $a/M$:
    \begin{align}
    \omega_{\ell mn}=\omega_{\ell mn}^{(0,0)}+\frac{a}{M}\omega_{\ell mn}^{(1,0)}+\frac{a^2}{M^2}\omega_{\ell mn}^{(2,0)}+\frac{\hat{\alpha}^2a^2}{M^2}\omega_{\ell mn}^{(2,2)}+{\cal O}(a^3/M^3)\;, \label{omega_series_exp}
    \end{align}
where $\omega_{\ell mn}^{(i,j)}$ denotes the coefficient of $(a/M)^i\hat{\alpha}^j$, and $n$ labels the overtone number.
The first three terms on the right-hand side correspond to the Kerr QNM frequency up to second order in $a/M$.
In particular, $\omega_{\ell mn}^{(0,0)}$ corresponds to the Schwarzschild QNM frequency.
By contrast, the fourth term, proportional to $\hat{\alpha}^2a^2/M^2$, originates from the parity-violating effects.
Note that, at least up to this order, the QNM frequencies are invariant under $\hat{\alpha}\to -\hat{\alpha}$.

One can then substitute Eq.~\eqref{omega_series_exp} into ${\cal A}(\omega;a,\hat{\alpha})=0$ and determine the coefficients~$\omega_{\ell mn}^{(i,j)}$ order by order.
The leading coefficient~$\omega_{\ell mn}^{(0,0)}$ is obtained by solving ${\cal A}(\omega;0,0)=0$, and the higher-order coefficients are then uniquely fixed provided that ${\cal A}_\omega \coloneqq \pa{\cal A}/\pa\omega \ne 0$ at $(\omega;a,\hat{\alpha})=(\omega_{\ell mn}^{(0,0)};0,0)$, i.e., that $\omega=\omega_{\ell mn}^{(0,0)}$ is a simple root.
Written explicitly, we have
    \begin{align}
    \begin{split}
    &\omega_{\ell mn}^{(1,0)}=-\frac{{\cal A}_a}{{\cal A}_\omega}\;, \qquad
    \omega_{\ell mn}^{(2,0)}
    =-\frac{{\cal A}_{aa}+2\omega_{\ell mn}^{(1,0)} {\cal A}_{\omega a} +\omega_{\ell mn}^{(1,0)}{}^2 {\cal A}_{\omega\omega}}{2{\cal A}_\omega}\;, 
    \\
    &\omega_{\ell mn}^{(2,2)}
    =-\frac{{\cal A}_{aa\hat{\alpha}\hat{\alpha}}+2\omega_{\ell mn}^{(2,0)}{\cal A}_{\omega\hat{\alpha}\hat{\alpha}}+2\omega_{\ell mn}^{(1,0)}{\cal A}_{\omega a\hat{\alpha}\hat{\alpha}}+\omega_{\ell mn}^{(1,0)}{}^2{\cal A}_{\omega\omega\hat{\alpha}\hat{\alpha}}}{4{\cal A}_{\omega}}\;,
    \end{split}\label{spec_coeff}
    \end{align}
where ${\cal A}$ with subscript(s) denotes the derivative of ${\cal A}$ with respect to the indicated variable(s), evaluated at $(\omega;a,\hat{\alpha})=(\omega_{\ell mn}^{(0,0)};0,0)$.
Here, we have used the fact that ${\cal A}_{\hat{\alpha}}=0$, which follows from the observation that $\hat{\alpha}$ enters only quadratically at leading order in the parity-violating effects.
Note also that, for modes with $m=0$, we have ${\cal A}_{a}=0$, and consequently $\omega_{\ell mn}^{(1,0)}=0$ in this case.

\subsection{\texorpdfstring{Small $\hat{\alpha}$}{Small alpha}}\label{ssec:small_coupling_constant}

In this section, we study the scalar QNM frequencies in the small-$\hat{\alpha}$ regime, allowing for arbitrary BH spin, from the non-rotating case to the near-extremal regime.
As discussed below Eq.~\eqref{QNM_BC}, the procedure employed in the previous section becomes inefficient away from the low-spin regime, since infinitely many modes with different multipole indices are coupled.
We therefore adopt a more robust and systematic approach, namely the spectral method~\cite{Chung:2023zdq,Chung:2023wkd}.

As in the previous section, we start from the modified Klein-Gordon equation~\eqref{KG_mod} for the scalar field~$\tilde{\phi}$.
Following Refs.~\cite{Chung:2023zdq,Chung:2023wkd}, we express the scalar field as
    \begin{align}
    \tilde{\phi}=\int{\rm d}\omega\sum_{m}{\rm e}^{-{\rm i}\omega t+{\rm i}m\varphi}A(r)\psi(r,\theta)\;, \label{eq:scalar_decomp}
    \end{align}
where $\psi(r,\theta)$ is a function that is regular on the domain~$[r_{\rm H},+\infty)\times[0,\pi]$.
The radial prefactor~$A(r)$ is chosen as\footnote{As discussed in Refs.~\cite{Chung:2023zdq,Chung:2023wkd}, one may in general include additional powers of $r$ and $r-r_{\rm H}$ in $A(r)$ to ensure regularity of $\psi(r,\theta)$ over the domain of interest; however, this is not necessary in the present analysis.}
    \begin{align}
    A(r)={\rm e}^{{\rm i}\omega r}r^{2{\rm i}M\omega}\bra{\fr{r-r_{\rm H}}{r}}^{-{\rm i}M\fr{r_{\rm H}}{r_{\rm H}-M}(\omega-m\omega_{\rm H})}\;, \label{A(r)}
    \end{align}
which satisfies the following boundary conditions, analogous to those in Eq.~\eqref{QNM_BC}:
    \begin{align}
    A(r)\sim
    \left\{\begin{array}{ll}
    {\rm e}^{-{\rm i}(\omega-m\omega_{\rm H})r_*} \;, & \qquad r\to r_\mathrm{H}\;, \\
    {\rm e}^{{\rm i}\omega r_*} \;, & \qquad r\to \infty\;.
    \end{array}\right.
    \end{align}

In order to adopt the spectral method, we now perform a coordinate transformation~$(r,\theta)\to (z,\chi)$ defined by
\begin{align}\label{eq:z_chi}
    z = \frac{2r_\mathrm{H}}{r} - 1\;, \qquad
    \chi = \cos{\theta}\;,
\end{align}
or, equivalently, 
\begin{align}
    r = \frac{2r_\mathrm{H}}{1 + z}\;, \qquad
    \theta = \arccos{\chi}\;,
\end{align}
so that the domain of the function~$\psi$ becomes $(z,\chi)\in[-1,1]\times[-1,1]$.
For notational simplicity, we use the same symbol~$\psi$ to denote the function expressed in the $(z,\chi)$ coordinates.
One can then expand the function~$\psi(z,\chi)$ in terms of the Chebyshev polynomials~$T_k(z)$ and associated Legendre polynomials~$P_l^{|m|}(\chi)$ as
\begin{align}
    \psi(z,\chi) = \sum_{k=0}^{\infty} \, \sum_{l=|m|}^{\infty}
    v_{kl} \, T_k(z) \, P_l^{|m|}(\chi)\;,
    \label{eq:psi_decomp}
\end{align}
where $v_{kl}$ are expansion coefficients.\footnote{As pointed out in Ref.~\cite{Chung:2023zdq}, one could in principle choose a different basis of functions to perform a spectral decomposition analogous to Eq.~\eqref{eq:psi_decomp}, provided that the basis is complete and orthogonal; the results should then be independent of the particular choice of basis.
Similarly, one could adopt a different coordinate transformation~$(r,\theta)\to (z,\chi)$.
For example, one may consider a one-parameter family of coordinate transformations~$r\to z$ defined by
    \begin{align}
    r=\frac{2 r_\mathrm{H}}{1 + z}\,\gamma + r_\mathrm{H}(1-\gamma)\;, \nonumber
    \end{align}
where $\gamma$ is a positive parameter, and the transformation reduces to our chosen coordinate mapping when $\gamma=1$.
One could then check the stability of the results against the choice of coordinate transformation and basis for the spectral decomposition, but this is beyond the scope of this work.}
Substituting Eqs.~\eqref{eq:scalar_decomp}, \eqref{A(r)}, and \eqref{eq:psi_decomp} into the modified Klein-Gordon equation~\eqref{KG_mod} with $\mu^2=0$, we obtain an equation involving $T_k(z)$, $P_l^{|m|}(\chi)$, and their derivatives up to second order.
These derivatives can be eliminated using the following formulae:
\begin{align}
\begin{split}
    \fr{\D^2 T_k(z)}{\D z^2} &= \frac{1}{1 - z^2}\left[z \fr{\D T_k(z)}{\D z} - k^2 T_k(z) \right]\;, \\
    \fr{\D T_k(z)}{\D z} &= \frac{k}{1 - z^2} \left[-z T_k(z) + T_{k-1}(z) \right]\;, \\
    \fr{\D^2 P_l^{|m|}(\chi)}{\D\chi^2} &= \frac{1}{1 - \chi^2} \left[2\chi \fr{\D P_l^{|m|}(\chi)}{\D\chi} - l(l + 1) P_l^{|m|}(\chi) + \frac{m^2}{1 - \chi^2} P_l^{|m|}(\chi) \right]\;.
\end{split}
\end{align}
Note in passing that eliminating the second derivative of $P_l^{|m|}(\chi)$ simultaneously removes the first derivative.
After removing all derivatives, we arrive at an algebraic equation of the form,
\begin{align}\label{eq:EoM_K_T_P}
    \sum_{k,k'=0}^{\infty} \, \sum_{l,l'=|m|}^{\infty}K_{kl k'l'}(z,\chi) \,
    v_{kl} \, T_{k'}(z) \, P_{l'}^{|m|}(\chi) = 0\;,
\end{align}
where 
$K_{kl k'l'}(z,\chi)$ are functions of $z$ and $\chi$.
In our setup, these functions are rational in $z$ and $\chi$, but by multiplying the equation by a suitable common denominator they can be taken to be polynomials in $z$ and $\chi$.
This allows us to re-expand the left-hand side in terms of the basis functions~$T_k(z)$ and $P_l^{|m|}(\chi)$ as
\begin{align}
    \sum_{k,k'=0}^{\infty} \, \sum_{l,l'=|m|}^{\infty}D_{kl k'l'} \,
    v_{kl} \, T_{k'}(z) \, P_{l'}^{|m|}(\chi) = 0\;,
\end{align}
where $D_{kl k'l'}$ are constants.
Practically, we consider only a finite number of QNMs, and therefore it is reasonable to truncate the infinite sums at finite orders, denoted by ${\cal N}_z-1$ and ${\cal N}_\chi+|m|-1$ for the sums over $k,k'$ and $l,l'$, respectively.
Note that ${\cal N}_z$ and ${\cal N}_\chi$ correspond to the numbers of Chebyshev polynomials and associated Legendre polynomials included in the spectral decomposition, respectively.
With this truncation, the orthogonality of the basis functions $T_k(z)$ and $P_l^{|m|}(\chi)$ implies
\begin{align}
    \sum_{k=0}^{{\cal N}_z-1} \, \sum_{l=|m|}^{{\cal N}_\chi+|m|-1}D_{kl k'l'} \,
    v_{kl} = 0\;.
\end{align}
Equivalently, this system can be written in matrix form as $\mathbb{D}\mathbf{v}=0$, where $\mathbf{v}$ is a vector with ${\cal N} \coloneqq {\cal N}_z \times {\cal N}_\chi$ components~$v_{kl}$, and $\mathbb{D}$ is an ${\cal N}\times{\cal N}$ matrix.
Since the matrix~$\mathbb{D}$ depends on the frequency~$\omega$ up to quadratic order, reflecting the fact that the modified Klein-Gordon equation~\eqref{KG_mod} contains time derivatives up to second order, the system can be written as
\begin{align}\label{eq:eigen_matrix_D}
    (M^2\omega^2 \mathbb{D}_2 + M\omega \mathbb{D}_1 + \mathbb{D}_0) \mathbf{v} = 0 \;,
\end{align}
where the matrices~$\mathbb{D}_2$, $\mathbb{D}_1$, and $\mathbb{D}_0$ are independent of $\omega$.
Note in passing that the BH mass parameter~$M$ has been introduced so that the frequency appears in the dimensionless combination~$M\omega$.
The matrix~$\mathbb{D}_2$ is invertible in general, and hence Eq.~\eqref{eq:eigen_matrix_D} can be recast into the form,
\begin{align}\label{eq:eigen_C_matrix}
    (M^2\omega^2 \mathbb{I}_{\cal N} + M\omega \mathbb{C}_1 + \mathbb{C}_0) \mathbf{v} = 0\;,
\end{align}
where $\mathbb{I}_{\cal N}$ denotes the ${\cal N}\times {\cal N}$ identity matrix, and we have defined $\mathbb{C}_1 \coloneqq \mathbb{D}_2^{-1} \mathbb{D}_1$ and $\mathbb{C}_0 \coloneqq \mathbb{D}_2^{-1} \mathbb{D}_0$.
Furthermore, by introducing a $2{\cal N}$-dimensional vector~$\mathbf{x}$ defined as
\begin{align}
    \mathbf{x} = \begin{pmatrix} \mathbf{v} \\ M\omega \mathbf{v} \end{pmatrix} \;,
\end{align}
the system can be rewritten as a standard eigenvalue problem,
\begin{align}\label{eq:Q_eigen}
    \mathbb{Q} \mathbf{x} = M\omega \mathbf{x}\;, \qquad
    \mathbb{Q} \coloneqq
    \begin{pmatrix}
    0 & \mathbb{I}_{\cal N} \\
    -\mathbb{C}_0 & -\mathbb{C}_1
    \end{pmatrix}\;.
\end{align}
Once the BH spin~$a$ (in units of the mass parameter~$M$), the parameter~$\hat{\alpha}$, and the azimuthal number~$m$ are specified, the matrices~$\mathbb{C}_0$ and $\mathbb{C}_1$ are determined.
The dimensionless scalar QNM frequencies~$M\omega$ are then obtained by solving the eigenvalue equation~\eqref{eq:Q_eigen}, which can be carried out numerically using standard linear algebra routines.

Before closing this section, several caveats are in order.
First, while QNMs in the standard Kerr case can be unambiguously labeled by $(\ell,m,n)$, it is nontrivial to define such labels in the conformal Kerr case with $\hat{\alpha}\neq 0$, since separation of variables in terms of the spheroidal harmonics~$S_{\ell m}(\theta)$ is no longer available.
Nevertheless, as we assume that the parity-violating effects are perturbative, each QNM of interest admits a smooth limit as $\hat{\alpha}\to 0$, corresponding to a Kerr QNM.
We therefore label the QNMs in the conformal Kerr case by $(\ell,m,n)$ according to their Kerr counterparts.
Second, not all eigenvalues of Eq.~\eqref{eq:Q_eigen} correspond to physical QNMs.
In practice, truncation of the spectral decomposition introduces spurious modes that contaminate the QNM spectrum.
It is thus necessary to choose ${\cal N}_z$ and ${\cal N}_\chi$ so as to suppress such contamination, while keeping them sufficiently small to ensure reasonable computational cost (see Sec.~\ref{ssec:consistency}).

\section{Results}\label{sec:res}

In this section, we present our results for the scalar QNM frequencies of the conformal Kerr background.
As a numerical consistency check of the spectral method, we first compute the standard Kerr QNM frequencies using this method in Sec.~\ref{ssec:consistency}.
We then discuss the QNM frequencies on the conformal Kerr background in the low-spin regime in Sec.~\ref{ssec:QNMs_low_spin}, and those including the high-spin regime (with small $\hat{\alpha}$) in Sec.~\ref{ssec:QNMs_small_alpha}.

\subsection{Numerical consistency check of spectral method}\label{ssec:consistency}

As briefly mentioned in Sec.~\ref{ssec:small_coupling_constant}, within the spectral method the QNM frequencies are obtained by solving the eigenvalue equation~\eqref{eq:Q_eigen}, but the resulting spectrum generally contains spurious modes.
Here, by spurious modes we mean eigenmodes that do not converge to physical QNMs with increasing truncation order.
Such modes are not artifacts of insufficient resolution of physical eigenfunctions, but arise from the truncation of an infinite-dimensional mode-coupling problem inherent in the spectral discretization.
For a general discussion of spurious eigenvalues in spectral methods, see, e.g., Sec.~7.6 of Ref.~\cite{Boyd2001}.

To mitigate such contamination, it is necessary to choose the truncation orders~${\cal N}_z$ and ${\cal N}_\chi$ appropriately, so as to suppress spurious modes while keeping the computational cost manageable.
For simplicity, we set ${\cal N}_z={\cal N}_\chi\eqqcolon N$ in what follows.
Before computing the QNM frequencies on the conformal Kerr background, we therefore analyze the Kerr (or Schwarzschild) case for several values of $N$, comparing our results with known Kerr QNM frequencies in the literature and examining how spurious modes are distributed relative to the physical QNM spectrum.

  \begin{figure}[t]
    \begin{center}
        \begin{minipage}[b]{0.49\textwidth}
        \begin{tabular}{rr}
        \includegraphics[keepaspectratio=true,width=88mm]{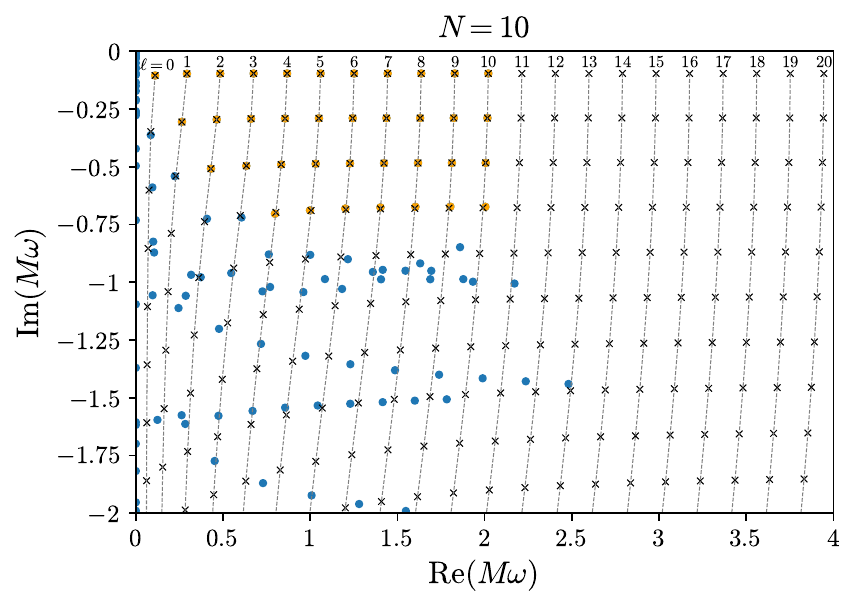}\\
        \includegraphics[keepaspectratio=true,width=88mm]{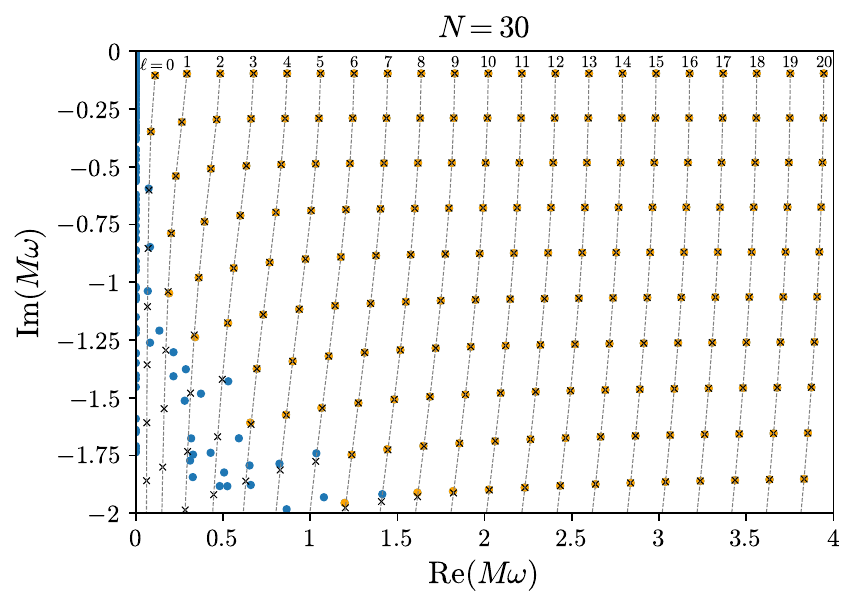}
        \end{tabular}
        \end{minipage}
        \begin{minipage}[b]{0.49\textwidth}
        \begin{tabular}{rr}
        \includegraphics[keepaspectratio=true,width=88mm]{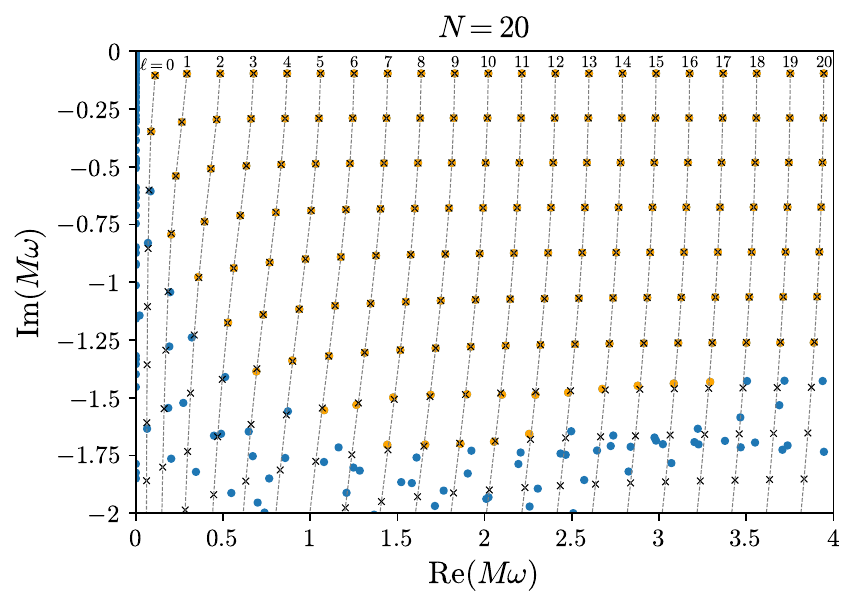}\\
        \includegraphics[keepaspectratio=true,width=88mm]{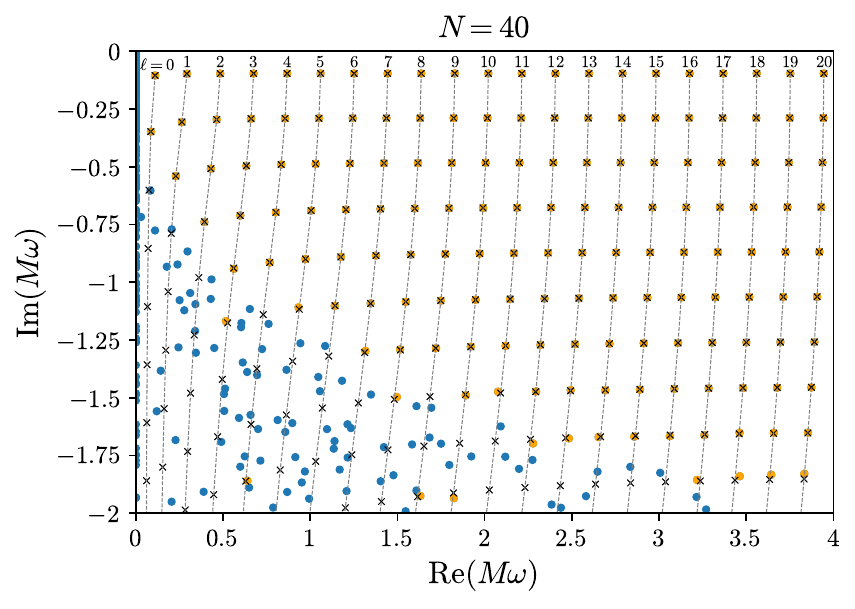}
        \end{tabular}
        \end{minipage}
       \caption{
       Scalar QNM frequencies for Schwarzschild BHs computed using the spectral method for $N\in\{10,20,30,40\}$.
       The value of $N$ used is indicated at the top of each panel.
       Cross markers denote the reference frequencies obtained with \texttt{qnm.py}, with dashed lines connecting modes of the same $\ell$.
       Circular markers show the solutions of the eigenvalue equation~\eqref{eq:Q_eigen}.
       Orange points agree with the reference values within a $1\%$ accuracy threshold and therefore correspond to physical QNMs, while blue points do not and are thus identified as spurious modes.
       }
       \label{fig:QNMF_for_various_N_at_a=0}
    \end{center}
  \end{figure}
  \begin{figure}[t]
    \begin{center}
        \begin{minipage}[b]{0.49\textwidth}
        \begin{tabular}{rr}
        \includegraphics[keepaspectratio=true,width=88mm]{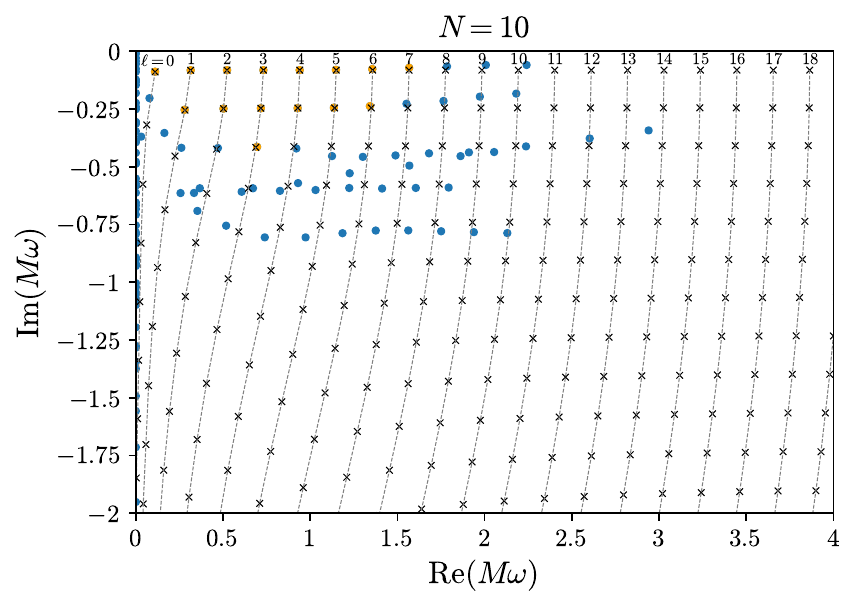}\\
        \includegraphics[keepaspectratio=true,width=88mm]{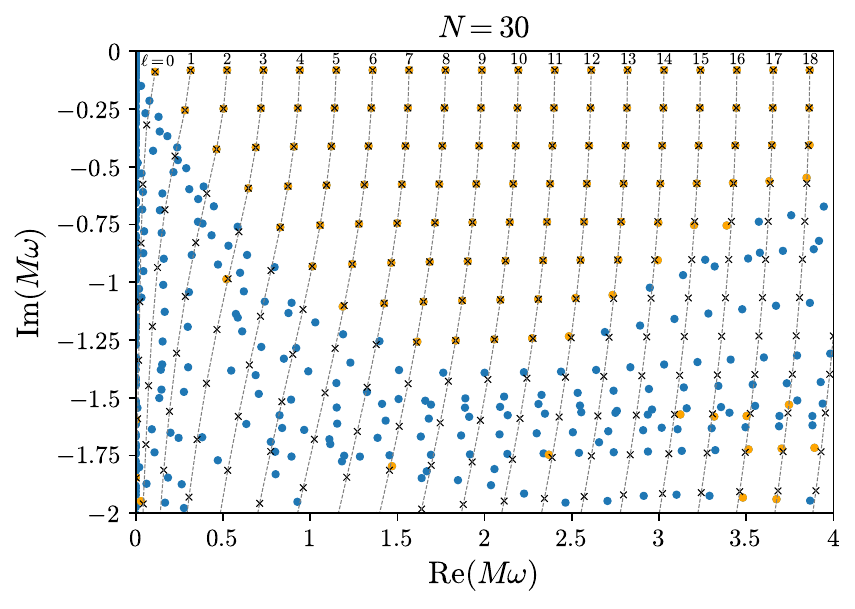}
        \end{tabular}
        \end{minipage}
        \begin{minipage}[b]{0.49\textwidth}
        \begin{tabular}{rr}
        \includegraphics[keepaspectratio=true,width=88mm]{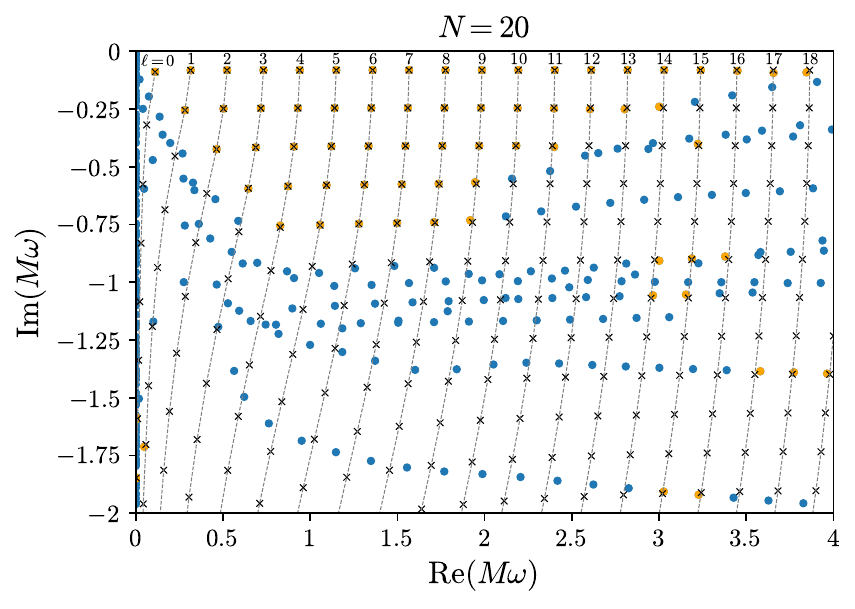}\\
        \includegraphics[keepaspectratio=true,width=88mm]{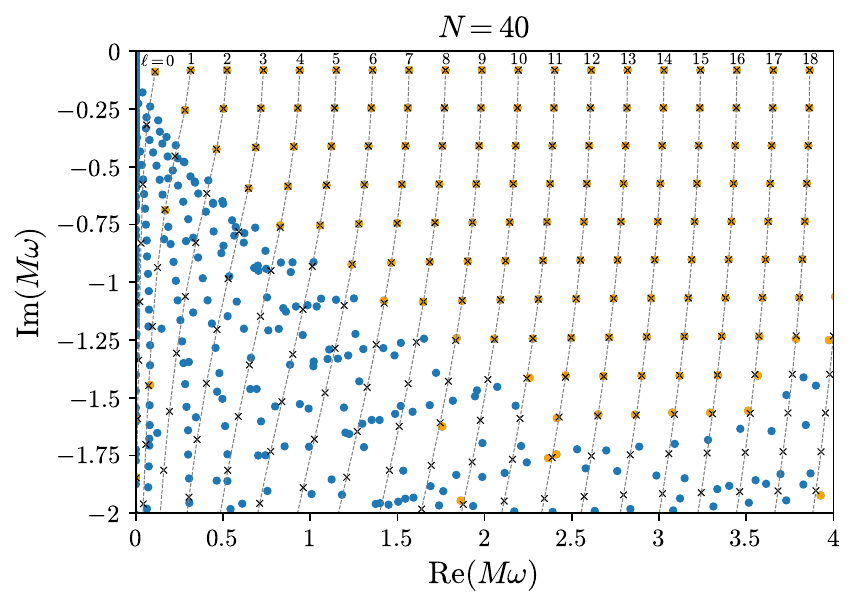}
        \end{tabular}
        \end{minipage}
       \caption{
       Scalar QNM frequencies for Kerr BHs with $a/M=0.99$ computed using the spectral method for $N\in\{10,20,30,40\}$.
       The value of $N$ used is indicated at the top of each panel.
       Here we consider the case~$m = 0$.
       Cross markers denote the reference frequencies obtained with \texttt{qnm.py}, with dashed lines connecting modes of the same $\ell$.
       Circular markers show the solutions of the eigenvalue equation~\eqref{eq:Q_eigen}.
       Orange points agree with the reference values within a $1\%$ accuracy threshold and therefore correspond to physical QNMs, while blue points do not and are thus identified as spurious modes.
       }
       \label{fig:QNMF_for_various_N_at_a=0.99}
    \end{center}
  \end{figure}

Let us first compute the scalar QNM frequencies of the Schwarzschild BH, denoted by $\omega_{\rm Sch}$, using the spectral method with $N\in \{10,20,30,40\}$.
Due to the spherical symmetry of the Schwarzschild background, $\omega_{\rm Sch}$ is independent of the azimuthal number~$m$.
The results are shown in Fig.~\ref{fig:QNMF_for_various_N_at_a=0}, where solutions of the eigenvalue equation~\eqref{eq:Q_eigen} are represented by orange or blue solid circles.
We focus on the region~$\textrm{Re}(M\omega)\in[0,4]$ and $\textrm{Im}(M\omega)\in[-2,0]$.
In all panels of Fig.~\ref{fig:QNMF_for_various_N_at_a=0}, the cross markers indicate reference values for the scalar Schwarzschild QNM frequencies, and the dashed lines connect modes with the same $\ell$.
Within each family, the uppermost points correspond to the fundamental mode ($n=0$), while modes with larger $|\textrm{Im}(M\omega)|$ correspond to higher overtones.
Most of these reference values are obtained independently using the Python package~\texttt{qnm.py}~\cite{Stein:2019mop};
however, for several higher overtones that are not supported by \texttt{qnm.py}, the reference frequencies are instead computed using a high-precision Julia code based on the improved Leaver-Nollert method developed in Ref.~\cite{Motohashi:2024fwt}.

The coloring of the eigenvalues obtained by the spectral method is determined by comparison with the reference values:
the orange data points agree with the reference frequencies at the $1\%$ level (chosen as a representative threshold), whereas the blue data points do not.
We therefore interpret the orange and blue data points as corresponding to physical QNMs and spurious modes, respectively.\footnote{One sometimes finds orange data points surrounded by blue points.
This is presumably coincidental, and it is therefore safer not to identify such isolated orange points as physical QNMs.
Note also that physical QNMs are stable under variations of the truncation order~$N$, whereas spurious modes are not.
Therefore, physical QNMs may be identified by examining the stability of each mode under variations of $N$, even without comparison with reference values.}
In Fig.~\ref{fig:QNMF_for_various_N_at_a=0}, the region corresponding to physical QNMs expands as the truncation order increases from $N=10$ to $N=30$.\footnote{For $N=10$, the QNMs with $\ell\le 10$ can be identified, whereas those with $\ell>10$ cannot. This is expected since, in the Schwarzschild limit ($a=0$), the index~$\ell$ of the spheroidal harmonics~$S_{\ell m}(\theta)$ coincides with the index~$l$ of the associated Legendre polynomials~$P_l^{|m|}(\cos\theta)$, and the spectral expansion is truncated at $l=10$.}
Although $N=40$ is the largest truncation order considered, the region of physical QNMs shrinks relative to that for $N=30$.
These observations suggest that an optimal choice of the truncation order is $N\simeq 30$.

Let us next compute the scalar QNM frequencies of the Kerr BH, denoted by $\omega_{\rm Kerr}$, using the spectral method with $N\in \{10,20,30,40\}$.
It should be noted that $\omega_{\rm Kerr}$ depends on the azimuthal number~$m$, and here we focus on the case $m=0$ for demonstration purposes.
For concreteness, we consider $a/M=0.99$.
The results are shown in Fig.~\ref{fig:QNMF_for_various_N_at_a=0.99}, where the meaning of the symbols and colors is the same as in Fig.~\ref{fig:QNMF_for_various_N_at_a=0}.
As in Fig.~\ref{fig:QNMF_for_various_N_at_a=0}, the region corresponding to physical QNMs expands as the truncation order increases from $N=10$ to $N=30$, while the region for $N=40$ is comparable to that for $N=30$.
This suggests that an optimal choice of the truncation order is around $N=30$ or $40$.
We also note that, for a fixed value of $N$, the region corresponding to physical QNMs is narrower than in the Schwarzschild case, making the identification of QNMs more challenging in the Kerr background.

The above results motivate us to use, e.g., $N=30$ when applying the spectral method to study the scalar QNM frequencies of the conformal Kerr BH.
Nevertheless, due to computational cost, we adopt $N=20$ in practice.
That said, in Secs.~\ref{ssec:QNMs_low_spin} and~\ref{ssec:QNMs_small_alpha}, we mainly focus on $\ell=0$ and $1$, for which $N=20$ performs comparably to $N=30$.

\subsection{QNMs for low spin}\label{ssec:QNMs_low_spin}

In this section, we investigate the scalar QNM frequencies of the conformal Kerr background in the low-spin regime, employing the matrix-valued Leaver's method and the spectral method described in Secs.~\ref{ssec:low_spin} and \ref{ssec:small_coupling_constant}, respectively.
As mentioned earlier, the analysis in Sec.~\ref{ssec:small_coupling_constant}, which is based on a small-$\hat{\alpha}$ expansion, is applicable to arbitrary BH spin.
We shall see that the results obtained from the two approaches are consistent when both $a$ and $\hat{\alpha}$ are small.

As argued in Sec.~\ref{ssec:low_spin}, in the low-spin regime the QNM frequencies admit the expansion~\eqref{omega_series_exp}.
When applying the matrix-valued Leaver's method, as explained around Eq.~\eqref{spec_coeff}, we expand the QNM-determining equation order by order in $a/M$ and $\hat{\alpha}$, and solve for the expansion coefficients analytically.
In the case of the spectral method, instead of performing an explicit order-by-order expansion, we compute the QNM frequencies for several values of $(a,\hat{\alpha})$ and extract the corresponding expansion coefficients by fitting the numerical results in the low-spin regime to Eq.~\eqref{omega_series_exp}.

\begin{table}[t]
\centering
\caption{Scalar QNM frequencies~$\omega_{\ell mn}$ of the conformal Kerr BH in the low-spin regime obtained by the matrix-valued Leaver's method or spectral method.
We list the expansion coefficients defined in Eq.~\eqref{omega_series_exp}, except for $\omega_{\ell mn}^{(1,0)}$ for $m=0$, as it vanishes for $m=0$.
For comparison, the corresponding Kerr reference values are also shown, except for $\omega_{\ell mn}^{(2,2)}$, which is absent in GR.}
\label{tab:low_spin_all}

\textbf{(a) $(\ell,m)=(0,0)$}\par\medskip
\begin{tabular}{c|cccc}
\toprule
$(\ell, m, n)$& Method & $M\omega_{\ell mn}^{(0,0)}$  
& $M\omega_{\ell mn}^{(2,0)}$ & $M\omega_{\ell mn}^{(2,2)}$ \\
\midrule
&Matrix-valued Leaver& ${0.110454 - 
0.104897{\rm i}}$ 
        & ${0.007725 + 0.009401{\rm i}}$ 
        & ${-0.056361- 
0.082739{\rm i}}$\\
$(0,0,0)$&Spectral& ${0.110444-0.104902{\rm i}}$ 
        & ${0.007802 + 0.009402{\rm i}}$ 
        & ${-0.056432 - 0.082795{\rm i}}$\\
&(Kerr reference values)& ${0.110455-0.104896{\rm i}}$ 
        & ${0.007 820 + 0.009 375{\rm i}}$ 
        & N/A \\
\midrule
&Matrix-valued Leaver& ${0.086048 - 
0.347993{\rm i}}$ 
        & ${0.013426 + 
0.052252{\rm i}}$ 
        & ${-1.142670 - 
0.785628{\rm i}}$\\
$(0,0,1)$&Spectral & ${0.088720-0.346663{\rm i}}$ 
        & ${0.013350 + 0.027241{\rm i}}$ 
        & ${-1.022060 - 0.735436{\rm i}}$\\
&(Kerr reference values)& ${0.086117-0.348052{\rm i}}$ 
        & ${0.008 768 + 0.038 676{\rm i}}$ 
        & N/A \\
\bottomrule
\end{tabular}

\medskip\bigskip

\textbf{(b) $(\ell,m)=(1,0)$}\par\medskip
\begin{tabular}{c|cccc}
\toprule
$(\ell, m,n)$ &Method & $M\omega_{\ell mn}^{(0,0)}$ & $M\omega_{\ell mn}^{(2,0)}$ & $M\omega_{\ell mn}^{(2,2)}$ \\
\midrule
&Matrix-valued Leaver& ${0.292936 - 
0.097660{\rm i}}$ 
        & ${0.019052 + 
0.008032 {\rm i}}$ 
        & ${0.011057 - 
0.002094{\rm i}}$\\
$(1,0,0)$&Spectral& ${0.292936-0.097660{\rm i}}$ 
        & ${0.019051 + 0.008033{\rm i}}$ 
        & ${0.011057 - 0.002094{\rm i}}$\\
&(Kerr reference values)& ${0.292936-0.097660{\rm i}}$ 
        & ${0.019051+0.008033{\rm i}}$ 
        & N/A \\
\midrule
&Matrix-valued Leaver& ${0.264449 - 
0.306257{\rm i}}$ 
        & ${0.028771 + 
0.030197{\rm i}}$ 
        & ${0.074089 - 
0.025233{\rm i}}$\\
$(1,0,1)$&Spectral& ${0.264439-0.306263{\rm i}}$ 
        & ${0.028638 + 0.030099{\rm i}}$ 
        & ${0.074102 - 0.025246{\rm i}}$\\
&(Kerr reference values)& ${0.264449-0.306257{\rm i}}$ 
        & ${0.028 658 + 0.030 066{\rm i}}$ 
        & N/A \\
\midrule
&Matrix-valued Leaver& ${0.229540 - 
0.540137{\rm i}}$ 
        & ${0.028882 + 
0.060308{\rm i}}$ 
        & ${0.265279 - 
0.644314{\rm i}}$\\
$(1,0,2)$&Spectral& ${0.229574-0.539450{\rm i}}$ 
        & ${0.039977 + 0.061781{\rm i}}$ 
        & ${0.256872 - 0.645725{\rm i}}$\\
&(Kerr reference values)& ${0.229539-0.540133{\rm i}}$ 
        & ${0.036 897 + 0.062 412{\rm i}}$ 
        & N/A \\
\bottomrule
\end{tabular}

\medskip\bigskip

\textbf{(c) $(\ell,m)=(1,1)$}\par\medskip
\begin{tabular}{c|ccccc}
\toprule
$(\ell, m,n)$& Method & $M\omega_{\ell mn}^{(0,0)}$ & $M\omega_{\ell mn}^{(1,0)}$ 
& $M\omega_{\ell mn}^{(2,0)}$ & $M\omega_{\ell mn}^{(2,2)}$ \\
\midrule
&Matrix-valued Leaver& ${0.292936 - 
0.097660{\rm i}}$ 
        & ${0.077158 + 0.000336{\rm i}}$ 
        & ${0.036991 + 
0.007291{\rm i}}$ 
        & ${0.006200 - 
0.003725{\rm i}}$\\
$(1,1,0)$&Spectral 
        & ${0.292936-0.097660{\rm i}}$ 
        & ${0.077158 + 0.000336{\rm i}}$ 
        & ${0.036991 + 0.007291{\rm i}}$ 
        & ${0.006200 - 0.003725{\rm i}}$\\
&(Kerr reference values)& ${0.292936-0.097660{\rm i}}$ 
        & ${0.077158+0.000336{\rm i}}$ 
        & ${0.036 991 + 0.007 291{\rm i}}$ 
        & N/A \\
\midrule
&Matrix-valued Leaver
        & ${0.264449 - 0.306257{\rm i}}$ 
        & ${0.090199 + 0.009861{\rm i}}$ 
        & ${0.043637 + 0.025810{\rm i}}$ 
        & ${0.026820 - 0.069681{\rm i}}$\\
$(1,1,1)$&Spectral
        & ${0.264439-0.306263{\rm i}}$ 
        & ${0.090232 + 0.009863{\rm i}}$ 
        & ${0.043480 + 0.025736{\rm i}}$ 
        & ${0.026819 - 0.069712{\rm i}}$\\
&(Kerr reference values)
        & ${0.264449-0.306257{\rm i}}$ 
        & ${0.090 198 + 0.009 861{\rm i}}$ 
        & ${0.043 525 + 0.025 679{\rm i}}$ 
        & N/A \\
\midrule
&Matrix-valued Leaver
        & ${0.229540 - 
0.540137{\rm i}}$ 
        & ${0.099124 + 
0.033681{\rm i}}$ 
        & ${0.042834 + 
0.049409{\rm i}}$ 
        & ${-0.072360 - 
0.442785{\rm i}}$\\
$(1,1,2)$&Spectral
        & ${0.229574-0.539450{\rm i}}$ 
        & ${0.098698 + 0.032678{\rm i}}$ 
        & ${0.054383 + 0.051151{\rm i}}$ 
        & ${-0.07584 - 0.436894{\rm i}}$\\
&(Kerr reference values)
        & ${0.229539-0.540133{\rm i}}$ 
        & ${0.099 124 + 0.033 673{\rm i}}$ 
        & ${0.050 848 + 0.051 522 {\rm i}}$ 
        & N/A \\
\bottomrule
\end{tabular}
\end{table}

We present the low-spin expansion coefficients of the scalar QNM frequencies defined in Eq.~\eqref{omega_series_exp} for several representative modes.
Specifically, the frequencies~$\omega_{000}$ and $\omega_{001}$ are listed in Table~\ref{tab:low_spin_all}(a), $\omega_{100}$, $\omega_{101}$, and $\omega_{102}$ in Table~\ref{tab:low_spin_all}(b), and $\omega_{110}$, $\omega_{111}$, and $\omega_{112}$ in Table~\ref{tab:low_spin_all}(c).
Note in passing that the results for $(\ell,m)=(1,-1)$ are omitted, as they follow directly from $\omega_{11n}$: the signs of the expansion coefficients multiplying odd powers of $a/M$ are flipped relative to those of $\omega_{11n}$, while those multiplying even powers of $a/M$ remain unchanged.
In each table, we report the expansion coefficients computed using both the matrix-valued Leaver's method and the spectral method.
Among these coefficients, $\omega_{\ell mn}^{(0,0)}$, $\omega_{\ell mn}^{(1,0)}$, and $\omega_{\ell mn}^{(2,0)}$ can be directly compared with the corresponding coefficients in the low-spin expansion of the Kerr QNM frequencies with respect to $a/M$, which are also included in the tables as reference values (see the \hyperref[AppA]{Appendix} for details on how these reference values are obtained).
Due to numerical limitations, we omit the second overtone~$\omega_{002}$ in 
Table~\ref{tab:low_spin_all}(a).
In fact, when applying the spectral method, even in the Schwarzschild case, the QNM with $\ell=0$ and $n=2$ cannot be resolved within a $1\%$ accuracy at the truncation order adopted here, as illustrated in Fig.~\ref{fig:QNMF_for_various_N_at_a=0}.

For the fundamental modes ($n=0$), we find that all coefficients obtained from the two methods agree with each other and with the Kerr reference values to good accuracy across all tables, demonstrating the reliability of both approaches.
For the first overtones ($n=1$), some discrepancies appear in the Kerr-related coefficients~$\omega_{\ell mn}^{(2,0)}$, particularly for $(\ell,m)=(0,0)$, indicating that the accuracy is somewhat limited in this case.
Nevertheless, the parity-violating coefficients~$\omega_{\ell mn}^{(2,2)}$ obtained from the two methods are consistent with each other.
For $(\ell,m)=(1,0)$ and $(1,1)$, all coefficients agree to good accuracy.
For the second overtones ($n=2$), both methods yield mutually consistent results, although noticeable deviations from the Kerr reference values persist for $\omega_{\ell mn}^{(2,0)}$.
Interestingly, as far as we have investigated, the parity-violating coefficients~$\omega_{\ell mn}^{(2,2)}$ are comparable to, or even larger than, the corresponding Kerr-related coefficients~$\omega_{\ell mn}^{(2,0)}$ in magnitude, and could therefore lead to a non-negligible correction when $\hat{\alpha}=\mathcal{O}(1)$.
In addition, the coefficients~$\omega_{\ell mn}^{(2,2)}$ have a negative imaginary part, implying that the QNMs of the conformal Kerr BH tend to decay faster than those of the standard Kerr BH in GR.

\subsection{\texorpdfstring{QNMs for small $\hat{\alpha}$}{QNMs for small alpha}}\label{ssec:QNMs_small_alpha}

In this section, we compute the scalar QNM frequencies of the conformal Kerr BH including the high-spin regime up to $a/M=0.99$, employing the spectral method.
As mentioned earlier, we now assume that parity-violating effects can be treated perturbatively, and this assumption remains valid for a nearly extremal BH as long as $\hat{\alpha}$ is small.
For concreteness, we present the results for $\hat{\alpha}=0.01$ and $0.02$, in comparison with the standard Kerr case in GR.
(See also footnote~\ref{footnote2} for a discussion of the validity of this choice of $\hat{\alpha}$.)

\begin{figure}[htbp]
        \begin{minipage}[b]{0.49\textwidth}
        \begin{tabular}{rr}
        \includegraphics[keepaspectratio=true,width=88mm]{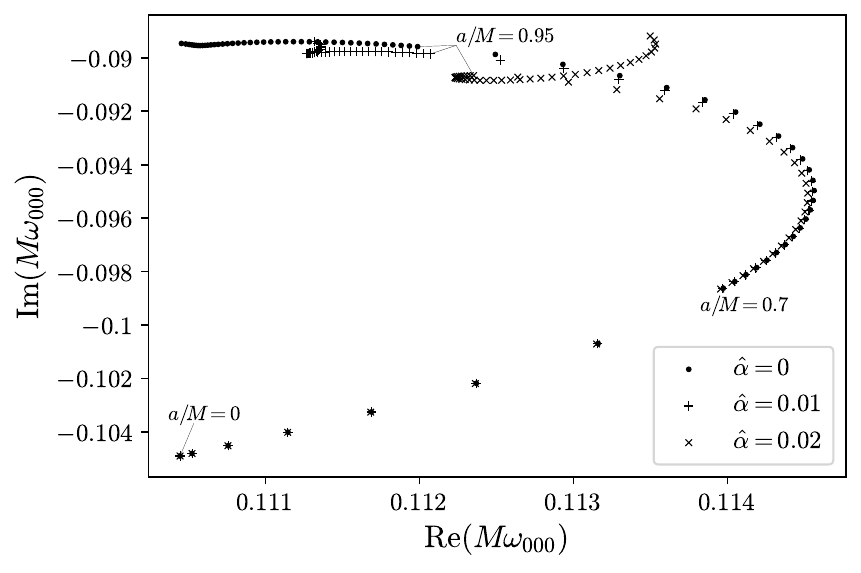}\\
        \includegraphics[keepaspectratio=true,width=88mm]{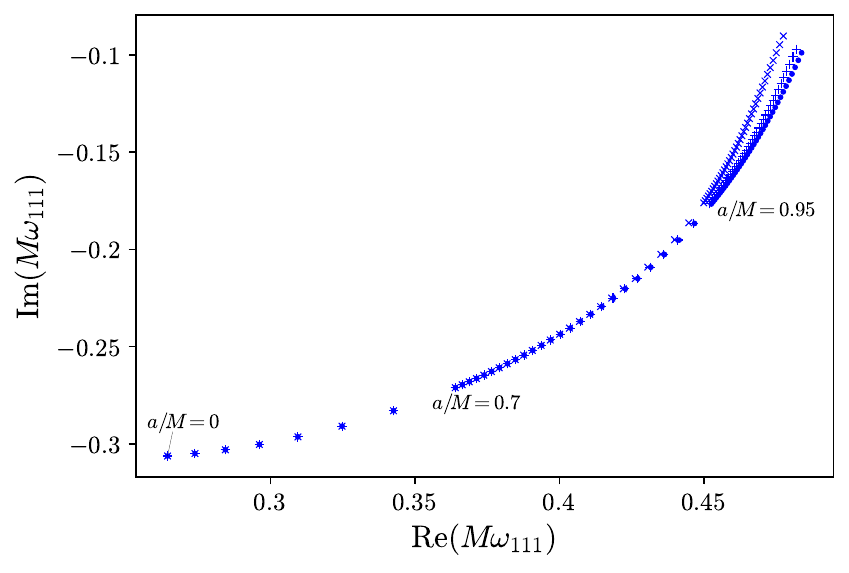}
        \end{tabular}
        \end{minipage}
        \begin{minipage}[b]{0.49\textwidth}
        \begin{tabular}{lr}
        \includegraphics[keepaspectratio=true,width=88mm]{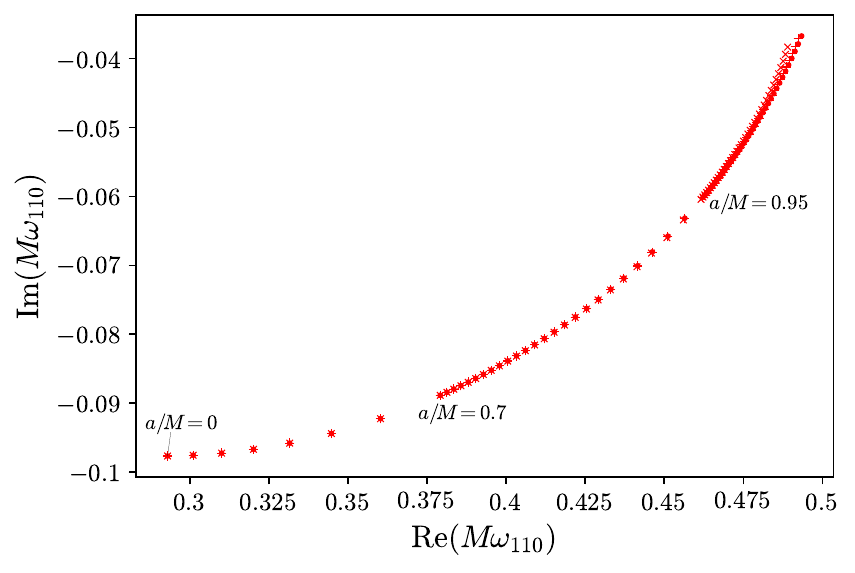}\\
        \includegraphics[keepaspectratio=true,width=88mm]{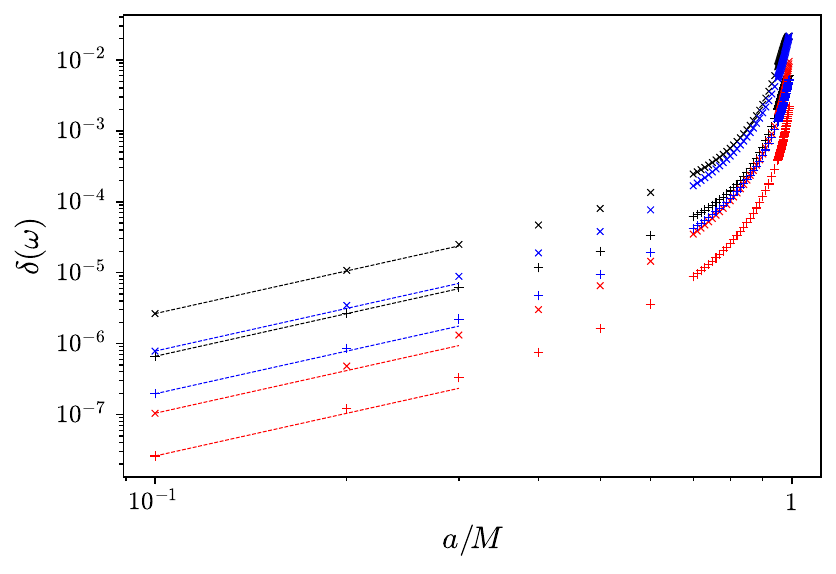}
        \end{tabular}
        \end{minipage}
  \caption{
        Scalar QNM frequencies for conformal Kerr BHs.
        We show the modes~$\omega_{000}$ (top left, black), $\omega_{110}$ (top right, red), and $\omega_{111}$ (bottom left, blue), computed as functions of the BH spin~$a/M$ for $\hat{\alpha}=0$ (solid circles), $0.01$ (plus symbols), and $0.02$ (crosses).
        The spin range~$a/M\in[0,0.99]$ is sampled with step sizes~$0.1$ for $a/M\in[0,0.7]$, $0.01$ for $a/M\in[0.7,0.95]$, and $0.001$ for $a/M\in[0.95,0.99]$.
        The bottom-right panel shows the relative deviation from the Kerr case, $\delta(\omega)\coloneqq|\omega/\omega_{\rm Kerr}-1|$, for $\hat{\alpha}=0.01$ and $0.02$, with colors matching those in the other panels and dashed lines indicating the analytic low-spin fits from Eq.~\eqref{omega_series_exp}.
       }
   \label{fig:qnms_high_spin}
  \end{figure}

In Fig.~\ref{fig:qnms_high_spin}, we plot the scalar QNM frequencies~$\omega_{000}$, $\omega_{110}$, and $\omega_{111}$ for conformal Kerr BHs in the spin range~$0\le a/M\le 0.99$.
In each panel, the frequencies are shown for $\hat{\alpha}=0$ (solid circles), $\hat{\alpha}=0.01$ (plus symbols), and $\hat{\alpha}=0.02$ (crosses), where the case~$\hat{\alpha}=0$ corresponds to the Kerr QNM frequencies.
At $a/M=0$ (the Schwarzschild limit), the solid circle, plus symbol, and cross mark coincide in each panel, as expected.
As $a/M$ increases, these data points separate and the deviations become sizable for $a/M\gtrsim0.7$, particularly in the near-extremal regime.

Interestingly, in the top-left panel of Fig.~\ref{fig:qnms_high_spin}, the mode~$\omega_{000}$ exhibits a turnover behavior in the near-extremal regime for $\hat{\alpha}=0.01$ and $0.02$, which is absent in the Kerr $\omega_{000}$ case.
A qualitatively similar turnover behavior is known to occur in the Kerr gravitational QNM $\omega_{225}$, where it arises from avoided crossings between different modes near an exceptional point and is accompanied by anomalous enhancement of the excitation factors~\cite{Motohashi:2024fwt}.
It is therefore tempting to speculate that the turnover behavior observed here may be related to similar non-Hermitian mode-coupling phenomena, and could potentially leave characteristic imprints in the time-domain response~\cite{Yang:2025dbn,PanossoMacedo:2025xnf}.
At present, it remains unclear whether the observed turnover of $\omega_{000}$ originates from an avoided crossing or instead reflects a generic spiraling trajectory of a single QNM in the complex-frequency plane.
Clarifying the underlying mechanism may require improved control over higher overtones and is left for future investigation.

The bottom-right panel of Fig.~\ref{fig:qnms_high_spin} displays the relative deviation from the Kerr case, defined as $\delta(\omega)\coloneqq|\omega/\omega_{\rm Kerr}-1|$, with $\omega_{\rm Kerr}$ denoting the Kerr QNM frequencies.
In the low-spin regime, fits based on the low-spin expansion~\eqref{omega_series_exp}, shown as dashed lines, are in good agreement with the numerical results for $a/M\lesssim0.3$.
More concretely, Eq.~\eqref{omega_series_exp} implies $\delta(\omega)\propto\hat{\alpha}^2 a^2/M^2$ when $a/M$ is small; accordingly, for each mode (indicated by different colors) and for a fixed value of $a$, the separation between the cross and plus data points is approximately controlled by the factor~$\hat{\alpha}^2$.
As the BH spin increases, the deviation grows rapidly, reaching values as large as $\delta(\omega)\sim 10^{-2}$ in the near-extremal regime, highlighting the enhanced impact of parity-violating effects on the scalar QNM spectrum.

\section{Conclusions}\label{sec:conc}

In this paper, we have investigated the effects of parity violation on quasinormal modes (QNMs) of black holes (BHs).
In particular, we have studied the QNMs of a test massless scalar field on a rotating BH solution in parity-violating gravity, known as the conformal Kerr BH.
This solution was obtained in Ref.~\cite{Tahara:2023pyg} via a conformal transformation of the Kerr solution in general relativity (GR), with the effects of parity violation encoded in the conformal factor.
As shown in Sec.~\ref{sec:QNM}, the scalar equation of motion on the conformal Kerr background can be recast as a Klein-Gordon equation with a spacetime-dependent effective mass term.
We have focused on regimes in which the parity-violating effects can be treated perturbatively, which applies when either the BH spin~$a$ (in units of the mass parameter~$M$) is small, or the parameter~$\hat{\alpha}$---characterizing the leading-order parity-violating effects---is small even for rapidly rotating BHs. 
We have analyzed the low-spin regime in Sec.~\ref{ssec:low_spin} and the small-$\hat{\alpha}$ regime in Sec.~\ref{ssec:small_coupling_constant}, noting that these two limits are not mutually exclusive.
In doing so, we have employed different but complementary methods for computing the QNMs, each suited to one of these two cases: the matrix-valued version of Leaver's method and the spectral method, respectively.

Our results for the scalar QNM frequencies on the conformal Kerr background are presented in Sec.~\ref{sec:res}.
As a numerical consistency check of the spectral method, we have computed the standard Kerr QNM frequencies using this method in Sec.~\ref{ssec:consistency}.
In Sec.~\ref{ssec:QNMs_low_spin}, we have focused on the low-spin regime and obtained a perturbative expansion for the QNM frequencies that includes the leading-order parity-violating corrections.
We have applied both the matrix-valued Leaver's method and the spectral method, and confirmed consistency between the results obtained using these two approaches.
In Sec.~\ref{ssec:QNMs_small_alpha}, we have focused on the small-$\hat{\alpha}$ regime.
Using the spectral method, we have computed the scalar QNM frequencies across a wide range of BH spins, from the non-rotating case to the near-extremal regime.
In particular, for near-extremal spins, we have found sizable deviations from the Kerr QNM frequencies, including an intriguing turnover behavior in the QNM spectrum.
Our results indicate that BH QNMs provide a useful probe of parity-violating physics in the strong-gravity regime.

There are several future directions that merit further investigation.
A natural extension of this work is to study the QNM spectrum of metric perturbations on the conformal Kerr background in the presence of matter fields.
The inclusion of matter is essential, since in vacuum the QNM spectrum remains identical to that in GR, reflecting the fact that the conformal transformation used to construct the conformal Kerr BH is an invertible redefinition of the metric.
Matter fields, however, distinguish between different frames through the metric to which they are minimally coupled, thereby rendering the physical predictions frame dependent.
It is also worthwhile to investigate how parity violation in the conformal Kerr background affects gravitational waveforms via a time-domain analysis.
In principle, such effects could lead to observable deviations from the GR predictions and may serve as a smoking-gun signature of gravitational parity violation.
Finally, from a technical perspective, it would be valuable to develop a mathematical understanding of how the distribution of spurious modes in the spectral method depends on the truncation order.
Such an understanding would help to more clearly distinguish non-physical modes from the physical QNM spectrum, thereby enhancing the practical applicability of the spectral method.
We leave these directions for future work.


\acknowledgments{
We thank Kei-ichiro Kubota for useful discussions.
This work was supported in part by JSPS (Japan Society for the Promotion of Science) KAKENHI Grant 
Nos.~JP22K03639 (H.M.) and JP23K13101 (K.T.),
and World Premier International Research Center Initiative (WPI), MEXT, Japan. 
V.Y.~is supported in part by grants for development of new faculty staff, Ratchadaphiseksomphot Fund, Chulalongkorn University and by the National Science, Research and Innovation Fund (NSRF) via the Program Management Unit for Human Resources \& Institutional Development, Research and Innovation Grant No.~B39G680009.
}


\appendix

\setcounter{equation}{0}
\renewcommand{\theequation}{A\arabic{equation}}

\makeatletter
\begingroup
\let\MakeTextUppercase\@firstofone
\let\MakeUppercase\@firstofone
\section*{Appendix: Low-spin expansion of Kerr QNM frequencies}\label{AppA}
\endgroup
\makeatother

In this appendix, we present the calculation of the Kerr QNM frequencies, which serves as a reference for comparison with the results obtained in the main text for the conformal Kerr BH.
Since our results in Sec.~\ref{ssec:QNMs_low_spin} are based on the low-spin expansion~\eqref{omega_series_exp} of the QNMs in the conformal Kerr spacetime, it is useful to perform an analogous expansion for the standard Kerr spacetime to clarify the effect of the conformal deformation.

Specifically, we compute the coefficients in the following low-spin expansion of the scalar QNM frequencies for Kerr BHs:
\be \label{eq:KerrQNMexp} \omega_\mathrm{Kerr} = \omega_\mathrm{Sch} + c_1\frac{a}{M} + c_2\frac{a^2}{M^2} + {\cal O}(a^3/M^3)\;, \ee
where $\omega_\mathrm{Sch}$ denotes the Schwarzschild QNM frequencies, $a$ is the BH spin parameter, and $M$ is the BH mass parameter.
Note in passing that Eq.~\eqref{eq:KerrQNMexp} corresponds to the first three terms in Eq.~\eqref{omega_series_exp}, with the identifications~$\omega_\mathrm{Sch}=\omega_{\ell mn}^{(0,0)}$, $c_1=\omega_{\ell mn}^{(1,0)}$, and $c_2=\omega_{\ell mn}^{(2,0)}$.

Using a high-precision Julia code based on the improved Leaver-Nollert method developed in Ref.~\cite{Motohashi:2024fwt}, we compute the scalar Schwarzschild QNM frequency~$M\omega_\mathrm{Sch}$ and the scalar Kerr QNM frequencies~$M\omega_k$ for $k=1,2,4$ evaluated at $a/M = k\delta$, with $\delta$ a small increment.
For $m=0$, from Eq.~\eqref{Ulm}, the linear coefficient vanishes ($c_1=0$), so only $c_2$ needs to be determined.
We obtain $c_2$ from
\be c_2 = \frac{\omega_1 - \omega_\mathrm{Sch}}{\delta^2}\;. \ee
To estimate the numerical error, we also evaluate $(\omega_2 - \omega_\mathrm{Sch}) / (2\delta)^2$ and compare the two results.
For high precision, the tolerance in the root-finding step of the Leaver-Nollert method was set to $10^{-40}$ and $\delta = 10^{-10}$, which is necessary to reliably extract $c_2$.
The relative error in $c_2$ then remains at ${\cal O}(10^{-20})$.

For $m\ne 0$, we determine $c_1$ and $c_2$ algebraically from the three data points~$(\omega_\mathrm{Sch}, \omega_1, \omega_2)$, separated by $\delta$, as
\be c_1 = -\frac{3 \omega_\mathrm{Sch} - 4 \omega_1 + \omega_2}{2 \delta}\;, \qquad 
c_2 = \frac{\omega_\mathrm{Sch} - 2 \omega_1 + \omega_2}{2 \delta^2}\;. \ee
We then repeat the procedure using $(\omega_\mathrm{Sch}, \omega_2, \omega_4)$ with spacing $2\delta$ to estimate relative errors.
For the $m\ne 0$ case, the tolerance in the root-finding was tightened to $10^{-60}$ and $\delta$ was set to $10^{-20}$.
The resulting relative errors of $c_1$ and $c_2$ are ${\cal O}(10^{-40})$ and ${\cal O}(10^{-20})$, respectively.

In this way, we obtain the Schwarzschild QNM frequencies and the coefficients of the low-spin expansion~\eqref{eq:KerrQNMexp} of the Kerr QNM frequencies up to ${\cal O}(a^2/M^2)$ with high numerical precision.
The results are summarized in Tables~\ref{tab:app1} and \ref{tab:app2}.
In Table~\ref{tab:app1}, the azimuthal number~$m$ is omitted, since the Schwarzschild QNM frequencies are independent of $m$ as a consequence of spherical symmetry.

\begin{table}
\caption{Scalar QNM frequencies for Schwarzschild BHs.}
\begin{tabular}{c|cc}
\toprule
$(\ell,n)$ & $\mathrm{Re}(M\omega_\mathrm{Sch})$ & $\mathrm{Im}(M\omega_\mathrm{Sch})$ \\
\midrule
$(0,0)$ & \num{0.11045493908041969} & \num{-0.10489571708688096} \\
$(0,1)$ & \num{0.08611691833639917} & \num{-0.34805244680646050} \\
$(0,2)$ & \num{0.07574193553517576} & \num{-0.60107859003580351} \\
$(0,3)$ & \num{0.07041013841746678} & \num{-0.85367731810553167} \\
\midrule
$(1,0)$ & \num{0.29293613326728271} & \num{-0.09765998891357822} \\
$(1,1)$ & \num{0.26444865060483254} & \num{-0.30625739155904712} \\
$(1,2)$ & \num{0.22953933493130167} & \num{-0.54013342501910721} \\
$(1,3)$ & \num{0.20325838618346365} & \num{-0.78829782278119803} \\
\bottomrule
\end{tabular}
\label{tab:app1}
\end{table}

\begin{table}
\caption{Coefficients of low-spin expansion~\eqref{eq:KerrQNMexp} of scalar QNM frequencies for Kerr BHs.}
\begin{tabular}{c|cccc}
\toprule
$(\ell,m,n)$ & $\mathrm{Re}(Mc_1)$ & $\mathrm{Im}(Mc_1)$ & $\mathrm{Re}(Mc_2)$ & $\mathrm{Im}(Mc_2)$  \\
\midrule
$(0,0,0)$ & \num{0} & \num{0} & \num{0.00782015710467309} & \num{0.00937481029605258} \\
$(0,0,1)$ & \num{0} & \num{0} & \num{0.00876793833110238} & \num{0.03867640697483717} \\
$(0,0,2)$ & \num{0} & \num{0} & \num{0.00111169609748111} & \num{0.06701569169258607} \\
$(0,0,3)$ & \num{0} & \num{0} & \num{-0.01061130477777911} & \num{0.09297415908467492} \\
\midrule
$(1,0,0)$ & \num{0} & \num{0} & \num{0.01905147787689066} & \num{0.00803283272099158} \\
$(1,0,1)$ & \num{0} & \num{0} & \num{0.02865827189954692} & \num{0.03006584519359344} \\
$(1,0,2)$ & \num{0} & \num{0} & \num{0.03689705463416485} & \num{0.06241192282380284} \\
$(1,0,3)$ & \num{0} & \num{0} & \num{0.03868897005632402} & \num{0.09775815503752841} \\
\midrule
$(1,\pm 1,0)$ & \num{\pm 0.07715782298934414} & \num{\pm 0.00033555891190736} & \num{0.03699084494601810} & \num{0.00729149801765656} \\
$(1,\pm 1,1)$ & \num{\pm 0.09019848430039377} & \num{\pm 0.00986058412915205} & \num{0.04352467725936047} & \num{0.02567910548973744} \\
$(1,\pm 1,2)$ & \num{\pm 0.09912369812164996} & \num{\pm 0.03367273841283092} & \num{0.05084782857360031} & \num{0.05152231422825025} \\
$(1,\pm 1,3)$ & \num{\pm 0.09762060474417335} & \num{\pm 0.05875502346681103} & \num{0.05343126953861450} & \num{0.08256349821893586} \\
\bottomrule
\end{tabular}
\label{tab:app2}
\end{table}


\bibliographystyle{mybibstyle}
\bibliography{bib}

\end{document}